\documentstyle[12pt]{article}

\newcommand{\p}[1]{\partial #1}
\newcommand{\dx}[1]{\int d^4 x #1}
\newcommand{\dy}[1]{\int d^4 x_1 #1}
\newcommand{\ddx}[1]{\int d^4 x_1 d^4 x_2 #1}
\newcommand{\dddx}[1]{\int d^4 x_1 d^4 x_2 d^4 x_3 #1}
\newcommand{\dk}[1]{\int {d^4 k \over (2\pi)^4 }#1}

\begin{document}
\begin{flushright}
NSC/USTC-29/96\\[10mm]
\end{flushright}
\begin{center}
{\Large{\bf Electromagnetic mass splittings of $\pi$, $a_1$,
 $K$, $K_1(1400)$ and $K^*(892)$}}\\[4mm]
Dao-Neng Gao\\
{\small Center for Fundamental Physics,
University of Science and Technology of China,\\
Hefei, Anhui,230026,P.R.China}\\[2mm]
Bing An Li\\
{\small Department of Physics and Astronomy,University of Kentucky,\\
Lexington,Kentucky 40506 USA}\\[2mm]
Mu-Lin Yan\\
{\small Chinese Center for Advanced Science and Technology(World Lab),\\
P.O.Box 8730, Beijing, 100080, P.R.China\\
Center for Fundamental Physics,
University of Science and Technology of China,\\
Hefei, Anhui,230026, P.R.China\footnote{Mailing address}}
\end{center}
\begin{abstract}
\noindent
To one-loop order and $O(\alpha_{em})$, the electromagnetic mass 
splittings of
$\pi$, $a_1$, $K$, $K_1(1400)$, and $K^*(892)$ are calculated in
the framework of $U(3)_L\times U(3)_R$ chiral
field theory. The logarithmic divergences emerging in the Feynman
integrations of the mesonic loops are factorized by using an
intrinsic parameter $g$ of this theory. No other additional parameters
or counterterms are introduced to absorb the mesonic loop divergences.
When $f_\pi$,$m_\rho$ and $m_a$ are taken as inputs,
the parameter $g$ will be determined and all the physical results
are finite and fixed.
Dashen's theorem is satisfied in the chiral SU(3) limit of this theory,
and a rather large violation of the theorem is revealed
at the order of $m_s$ or $m_K^2$.
Mass ratios of light quarks have been determined.
A new relation for electromagnetic corrections to masses of
axial-vector mesons is obtained. It could be regarded as a generalization of
Dashen's theorem. Comparing with data, it is found that the non-electromagnetic
mass difference of $K^*$ is in agreement with the estimation of Schechter,
Subbaraman,and Weigel.
\end{abstract}
\vskip 1cm
\section{Introduction}

Calculating the electromagnetic mass splittings of the low-lying mesons is
an important issue in non-perturbative quantum chromodynamics(NP-QCD).
This topic has intrigued particle physicists for many years[1--9].
Recently, a chiral field theory of pseudoscalar, axial-vector
and vector mesons(called as $U(3)_L\times U(3)_R$
chiral fields theory of mesons)has been proposed\cite{Li1,Li2}.
This theory can be regarded as a realization
of chiral symmetry, current algebra and vector meson dominance(VMD).
In this paper, we try to
present systematical calculations of electromagnetic masses of $\pi$, $a_1$, $K$, $K_1(1400)$ and
$K^*(892)$ in the framework of this theory.

It's well known that chiral perturbation theory($\chi$PT) is rigorous and
phenomenologically successful in
describing the physics of the pseudoscalar mesons at low
energies\cite{SGL}. The effective Lagrangian of $\chi$PT depends on ten
chiral coefficients which are determined by comparison with the experimental 
low-energy
information. Models attempt to extend the $\chi$PT to include
more low-lying mesons should predict these ten coefficients by fitting
data in $\chi$PT.
$U(3)_L\times U(3)_R$ chiral field theory has been studied at the tree
level\cite{Li2}, and the theoretical results
agree well with data. This theory has also been successfully applied to study
$\tau$ mesonic decays systematically\cite{Li31}. In Ref.\cite{Li4},
the ten coefficients of $\chi$PT have been predicted at 
about $\Lambda\sim$ 2GeV in this theory. The coefficients
of $\chi$PT are expressed by a universal coupling constant $g$ and the ratio
$\frac{f_\pi^2}{m_\rho^2}$ which have been fixed in\cite{Li1,Li2}.
The authors of Refs.\cite{DRV,EGLPR} have found that the vector meson dominates
the structure of the phenomenological chiral Lagrangian. Two of the coefficients
obtained in Ref.\cite{Li1} are the same as the ones in Ref.\cite{DRV}.
The relations $2(L_1+L_2)+L_3=0$ and $L_4^V=L_6^V=L_7^V=0$ found in
Ref.\cite{Li4} have already been obtained in Ref.\cite{EGLPR}.
A very small $L_8$ predicted in Ref.\cite{Li4} is not in contradiction
with $L_{8}^{V}=0$ found in Ref.\cite{EGLPR}. The expression of $L_9$
presented in Ref.\cite{Li4} is similar to the one obtained in
Ref.\cite{EGLPR}. When taking $g=1$, the $L_2^V=G^2_V/(16M^2_V)$ is the
same as the expression presented in Ref.\cite{Li4}.

In Ref.\cite{Wang}, starting from the $U(3)_L\times U(3)_R$ chiral fields
theory of mesons,
the authors use the path integration method to derive
$L_1, L_2, L_3, L_9$ and $L_{10}$. 
The results are in agreement with
the experimental values of the $L_i$ at $\mu=m_\rho$ in $\chi$PT. Therefore, 
the low-energy limit of this theory is indeed equivalent to $\chi$PT, and
the QCD constraints discussed
in Ref.\cite{EGLPR} are met by this theory.

$U(3)_L\times U(3)_R$ chiral fields theory of mesons provides a unified
description of meson physics at low energies.
VMD in the meson physics is a natural consequence of this theory instead of
an input. Therefore, the dynamics of the electromagnetic interactions
of mesons has been introduced and established naturally. On the other hand,
this theory starts with a chiral Lagrangian of quantum quark fields
within mesonic background fields, and the chiral dynamics for mesons
comes from the path integration over quark fields. A cut-off
$\Lambda$(or $g$ in Ref.\cite{Li1}) has to be introduced to absorb the logarithmic
divergences due to quark loops. Thus $g$(or $\Lambda$) will serve as an
intrinsic parameter in this truncated fields theory.
Therefore, it is legitimate to use the $g$ to factorize the logarithmic
divergences of loop diagrams in calculating the
electromagnetic mass splittings of the low-lying mesons\cite{LY}.

The basic Lagrangian of this chiral fields theory is(hereafter we use the
notations in Refs.{\cite{Li1,Li2})
\begin{eqnarray}
\lefteqn{{\cal L}=\bar{\psi}(x)(i\gamma\cdot\partial+\gamma\cdot v
+\gamma\cdot a\gamma_{5}
-mu(x))\psi(x)}\nonumber \\
& &+{1\over 2}m^{2}_{1}(\rho^{\mu}_{i}\rho_{\mu i}+
\omega^{\mu}\omega_{\mu}+a^{\mu}_{i}a_{\mu i}+f^{\mu}f_{\mu})\nonumber\\
& &+{1\over 2}m^2_2(K_{\mu}^{*a}K^{*a\mu}+K_1^{\mu}K_{1\mu})\nonumber \\
& &+\frac{1}{2}m^2_3(\phi_{\mu} \phi^{\mu}+f_s^{\mu}f_{s\mu})
\end{eqnarray}
with
\begin{eqnarray}
& &u(x)=exp[i\gamma_{5} (\tau_{i}\pi_{i}+\lambda_a K^a+\eta
+\eta^{\prime})],\nonumber\\
& &a_{\mu}=\tau_{i}a^{i}_{\mu}+\lambda_a K^a_{1\mu}+(\frac{2}{3}
+\frac{1}{\sqrt{3}}
\lambda_8)f_{\mu}+(\frac{1}{3}-\frac{1}{\sqrt{3}} \lambda_8)f_{s\mu},\nonumber\\
& &v_{\mu}=\tau_{i}\rho^{i}_{\mu}+\lambda_a K_{\mu}^{*a}+(\frac{2}{3}+
\frac{1}{\sqrt{3}}\lambda_8)\omega_{\mu}+(\frac{1}{3}-
\frac{1}{\sqrt{3}}\lambda_8)\phi_{\mu}.
\end{eqnarray}
where $i$=1,2,3 and $a$=4,5,6,7.
The $\psi$ in Eq.(1) is $u$,$d$,$s$ quark fields. $m$ is a parameter related to the quark condensate.
Here, the mesons are bound states in QCD,and they are not fundamental fields. Therefore,in Eq.(1) there are no
kinetic terms for these fields and the kinetic terms will be
generated from quark loops.  

According to Refs.\cite{Li1,Li2}, the effective Lagrangian ${\cal L}_{RE}$ and
${\cal L}_{IM}$ can be evaluated by performing the path integrations over quark
fields.
In order to absorb the logarithmic divergences in the effective Lagrangian,
as mentioned above,
it is necessary to introduce a universal coupling constant $g$
as follows
\begin{equation}
g^2={8\over 3}{ N_{c} \over (4\pi)^{D/2}} {D \over 4}
({\mu^2 \over m^2})^{\epsilon /2} \Gamma (2-{D\over 2})
=\frac{1}{6}\frac{F^2}{m^2}
\end{equation}
Also, following Refs.\cite{Li1,Li2}, after defining the physical meson-fields, we have
\begin{eqnarray}
m_{a}^2 & =&({1\over 1-{1\over 2\pi^2 g^2}})(m_{\rho}^2
     +{F^2 \over g^2} ), \\
m_{K_1}^2& =& ({1\over 1-{1\over 2\pi^2 g^2}})(m^2_{K^*}+{F^2 \over g^2}),
\end{eqnarray}
with 
\begin{eqnarray}
&&F^2={f_{\pi}^2 \over 1-{2c \over g}},\;\;\;
 c={f_{\pi}^2 \over 2gm_{\rho}^2},\\
&& F^2={f_k^2 \over 1-{2 c^{\prime}\over g}},\;\;\;
 c^{\prime}={f_k^2 \over 2 g m^2_{K^{*}}},\\
&& m^2=\frac{F^2}{6 g^2}.
\end{eqnarray}
Combining Eq.(4) with (6), and taking $f_\pi$,$m_\rho$,$m_a$ as inputs, the
parameter $g$ will be fixed.

VMD has been well established in studying electromagnetic
interactions of hadrons\cite{JS1}. To the present theory, the interactions
between photon and the vector meson fields of $\rho_0$,$\omega$ and $\phi$ can be found through following
substitutions\cite{Li2}
\begin{eqnarray}
\rho_\mu^3 \longrightarrow \rho_\mu^3 +
          {1 \over 2}eg A_\mu,  \\
\omega_\mu \longrightarrow \omega_\mu +
	       {1 \over 6} e g A_\mu,\\
\phi_\mu \longrightarrow \phi_\mu-{1\over 3 \sqrt{2}} e g A_\mu.
\end{eqnarray}
The $\rho^3$ (or $\rho^0$)-photon, $\omega$-photon and $\phi$-photon 
interaction Lagrangians are
\begin{eqnarray}
{\cal L}_{\rho \gamma}=-{1 \over 2}eg \partial_\mu
    \rho_\nu^3 (\partial^\mu A^\nu -\partial^\nu A^\mu), \\
{\cal L}_{\omega \gamma}=-{1 \over 6}eg \partial_\mu
    \omega_\nu (\partial^\mu A^\nu -\partial^\nu A^\mu), \\
{\cal L}_{\phi \gamma}={1\over 3 \sqrt{2}} eg \partial_\mu
    \phi_\nu (\partial^\mu A^\nu-\partial^\nu A^\mu).
\end{eqnarray}

Using ${\cal L}_i (\phi ,\gamma,...)|_{\phi=\pi,a,v}$ we can 
calculate the following S-matrix
\begin{equation}
S_{\phi}=\langle \phi |T{\rm exp}[i\int dx^4 {\cal L}_i
(\phi,\gamma,...)]-1|\phi\rangle |_{\phi=\pi,a,v}.
\end{equation}
On the other hand $S_\phi$ can also be expressed in terms of
the effective Lagrangian of $\phi$ as
$$
S_\phi=\langle \phi |i\dx {\cal L}_{{\rm eff}}(\phi)|\phi
\rangle.
$$
Noting ${\cal L}=\frac{1}{2}\partial_\mu\phi\partial^\mu\phi-
{1\over 2}m_{\phi}^2 \phi^2$,
then the electromagnetic interaction correction to the mass of $\phi$ reads
\begin{equation}
\delta m_\phi^2 ={2iS_\phi \over \langle \phi |\phi^2 |
\phi \rangle }.
\end{equation}
where $\langle \phi |\phi^2 |\phi \rangle =
\langle \phi |\dx \phi^2(x) |\phi \rangle $. 
We adopt dimensional regularization to do loop-calculations,
and use Eq.(3) to factorize the divergences. Thus, all of
virtual photon contributions to the masses of the low-lying  
mesons can be computed systematically and analytically.

The purposes of our investigations in this paper are threefold, which are stated
as follows.

1. We try to present a systematical method to derive the electromagnetic masses of the mesons
by employing $U(3)_L\times U(3)_R$ chiral fields theory.
Nearly thirty years ago, Das et al.\cite{DGMLY} obtained a finite result
of $\pi^{+} -\pi^{0}$ mass-difference by using current algebra techniques,
and especially relying on the second Weinberg's sum rule to cancel the
divergences in it (further investigations on it, see\cite{GPP,DHW}).
However, the second Weinberg's sum rule is not satisfied
experimentally\cite{JS2,ER,PDG}.
Actually, many people have found the existence of the
divergence in calculating the electromagnetic mass of $\pi-$mesons in the
effective chiral Lagrangian theories\cite{BB,BP,EGPR,UB}.
Especially, in Ref.\cite{BB}, when the corrections
of perturbative QCD to $m_{\pi^+}-m_{\pi^0}$ were investigated in a
chiral model, a dependence of $m_{\pi^+}-m_{\pi^0}$
on ultraviolet cutoff has been revealed. In Refs.\cite{EGPR,UB}, in order to
remove this divergence, the counterterms have been introduced.
These facts mean that we could not expect such a cancellation between
the divergent terms works without any additional assumptions,
in particular, when the strange-flavor mesons are involved.
The method in this paper is systematical, and the logarithmic divergences
from mesonic loops can be factorized by using the intrinsic parameter($g$ or
$\Lambda$) of the theory. There is no need of introducing other new parameter
or the counterterms to absorb the mesonic loop divergences.
The spirit of this method will be shown in Sec.2 by re-examining
the calculations of electromagnetic mass difference of charge and neutral $\pi-$mesons
in the present theory.

2. It is straightforward to extend our method to the studies of the electromagnetic
masses of strange-flavor mesons. The smallness of $u$,$d$ quark masses allows
the calculations in the chiral limit for non-strange mesons. However, the
large strange quark mass will bring a significant contribution to the
electromagnetic self-energies of the strange-flavor mesons.
Dashen's theorem\cite{DA} states that the square electromagnetic mass differences
between the charged pseudoscalar mesons and their corresponding neutral
partners are equal in the chiral SU(3) limit,i.e.,
$$
{(m^2_{K^+}-m^2_{K^0})_{EM}=(m^2_{\pi^+}-m^2_{\pi^0})_{EM}}
$$
The subscript EM denoted the electromagnetic mass.
The significant SU(3) symmetry breaking will lead to the violations
of this theorem. Furthermore, it has been known that the $\pi^+$--$\pi^0$ mass
difference is almost entirely electromagnetic in origin, however,
the contributions of the $K^+$--$K^0$ mass difference are from both
electromagnetic interactions and the $u$-$d$ quark mass
difference. Thus, it is of interest to calculate the
electromagnetic mass difference between $K^+$ and $K^0$ to the leading order in
quark mass expansion both to increase the understanding of the low-energy
dynamics and to aid in the extraction of current mass ratios of light quarks.
The latter reflects the breaking effect of isospin symmetry\cite{DHW,GTW,KM}.
Therefore, the quark mass term $-\bar{\psi}M\psi$($M$=diag$(m_u,m_d,m_s)$
is quark mass matrix), which represents the explicit chiral symmetry breaking
in the present theory, should be added into Eq.(1)
when the electromagnetic masses of the strange-flavor mesons are calculated.
The nonzero quark masses will yield the mass terms of pseudoscalar mesons
in addition to ${\cal L}_{RE}$(Explicit quark mass parameter do not occur
in the abnormal part effective Lagrangian). To the leading order in quark
mass expansion,the masses of the octet pseudoscalar mesons have been
derived in Ref.\cite{GOR}(Gell-Mann,Oakes, Renner formulas),which read
\begin{eqnarray}
m^{2}_{\pi^{+}}=m^2_{\pi^{0}}=-{2\over f^{2}_{\pi}}(m_{u}+m_{d})
\langle 0|\bar{\psi}\psi|0\rangle,\nonumber \\
m^{2}_{K^{+}}=-{2\over f^{2}_{k}}(m_{u}+m_{s})\langle 0|\bar{\psi}\psi|0\rangle,
\nonumber \\
m^{2}_{K^{0}}=-{2\over f^{2}_{k}}(m_{d}+m_{s})\langle 0|\bar{\psi}\psi|0\rangle,
\nonumber \\
m^{2}_{\eta}=-{2\over 3f^{2}_{\eta}}(m_{u}+m_{d}+4m_{s})
\langle 0|\bar{\psi}\psi|0\rangle.
\end{eqnarray}
where $\langle 0|\bar{\psi}\psi|0\rangle$ is the quark condensate of the light
flavors\cite{Li1,Li3}

3. All of the low-lying mesons including pseudoscalar, vector and axial-vector
mesons are involved in this theory. This makes it possible to evaluate the
electromagnetic masses of vector and axial-vector mesons
besides pseudoscalar $\pi$ and $K$.
The electromagnetic mass splittings of $a_1$ and $K_1(1400)$ are calculated,
and in the chiral SU(3) limit we obtain a new relation
$$(m_{a^+}^2-m_{a^0}^2)_{EM}=(m_{K_1^+}^2-m_{K_1^0}^2)_{EM}
$$
which could be regarded as a generalization of Dashen's theorem.
The electromagnetic masses of $K^*(892)$ are also derived. Using the
experimental value of $(m_{K^{*+}}-m_{K^{*0}})$, the non-electromagnetic mass
difference of $K^{*+}$ and $K^{*0}$ is estimated. The result is close to
the one given in Ref.\cite{SSW}.

The contents of this paper are organized as follows:
Sec.2, electromagnetic mass splitting of $\pi-$mesons.
Sec.3, electromagnetic mass splitting of $a_1-$mesons. Sec.4, we will extend this method
to the case of $K-$mesons, and give the violations of Dashen's
theorem at the leading order in quark mass expansion.
Sec.5, electromagnetic mass splitting of $K_1(1400)$. Sec.6, electromagnetic mass splitting
of $K^*(892)$. Sec.7, the discussion and summary
of the results.

\section{$\pi^+-\pi^0$ electromagnetic mass difference}

In this Section and next Section,we will restrict our calculations in
the two-flavor case because the strange quark has
no effect on the electromagnetic self-energies of pions and $a_1$ mesons,
and the smallness of $u$,$d$ quark masses allows the calculations
in the chiral limit.
Note that the contributions from ${\cal L}_{IM}$ are
proportional to $m_\pi^2$, which can be neglected in the chiral limit.
Thus, from ${\cal L}_{RE}$(Eq.(13) in Ref.\cite{Li1}),
the interaction Lagrangians
contributing to $\pi^+-\pi^0$ electromagnetic mass difference for massless
pions read
\begin{eqnarray}
{\cal L}_{\rho \rho \pi \pi}&=&{2F^2 \over g^2 f_\pi^2}\rho_\mu^i
  \rho^{j\mu}(\pi^2\delta_{ij}-\pi_i\pi_j)+{1 \over \pi^2 g^2 f_\pi^2}
  \p_\nu\rho_\mu^i \p^\nu\rho^{j\mu}
  (\pi^2\delta_{ij}-\pi_{i}\pi_{j}), \\
{\cal L}_{\rho \pi a}&=&-{2F^2\gamma \over f_\pi g^2}
  \rho_\mu^i \epsilon_{ijk}\pi_k a^{j\mu}
  +{\gamma \over f_\pi g^2 \pi^2} 
    \rho_\mu^i \epsilon_{ijk}\pi_k\p^2 a^{j\mu}, \\ 
{\cal L}_{\rho \pi \pi}&=&{2 \over g }
    \rho_\mu^i \epsilon_{ijk}\pi_k 
    (-\p^\mu \pi_j +{1 \over 2\pi^2 F^2}\p^2\p^\mu\pi_j). 
\end{eqnarray}
where $\gamma=(1-\frac{1}{2\pi^2 g^2}))^{-1/2}$.

Using VMD, i.e., the substitution (9),and (18)-(20) we get
all of corresponding photon-$\pi$ interaction Lagrangians
${\cal L}_{\gamma\gamma\pi\pi},{\cal L}_{\gamma\rho\pi\pi},
{\cal L}_{\gamma\pi a}$  and $ {\cal L}_{\gamma\pi\pi}.$
Combining them with ${\cal L}_{\rho\gamma} $ (Eq.(12)), we
can calculate $S_\pi$ (Eq.(15)), and obtain the $\pi^+-\pi^0$
mass difference due to electromagnetic interactions. The corresponding
Feynman diagrams are shown in Figures 1,2 and 3.
Denoting the corresponding $S$-matrices as $S_\pi (1)$,$S_\pi (2)$ and $
S_\pi (3)$ respectively, we have
$$
S_\pi =S_\pi (1)+S_\pi (2)+S_\pi (3).
$$
We  will compute them up to $O(e^2)$ separately below.
In order to show the gauge independence of the final results
explicitly, we take the most general linear gauge condition
for electromagnetic fields to all diagram calculations in
this paper. Namely, the $A_\mu$-propagator with an arbitrary
gauge parameter $a$ is taken to be
\begin{eqnarray}
{\Delta_{F}^{(\gamma)}} _{\mu\nu}(x-y)&=&\dk
     {{\Delta_{F}} ^{(\gamma)}} _{\mu\nu}
     (k)e^{-ik(x-y)}, \nonumber \\
{\Delta_F^{(\gamma)}} _{\mu\nu}(k)&=&
 {-i \over k^2} [g_{\mu\nu}-(1-a){k_\mu k_\nu \over k^2}].
\end{eqnarray}

Firstly, we compute $S_\pi (1)$ (Fig. 1). From Eqs.(15),(12),
(18) and (9), we have
\begin{eqnarray}
S_\pi (1) = {{\langle \pi |T [i\dy {\cal L}_{ \gamma \gamma \pi \pi}} 
( x_1 )
+{ i^2 \over  2! } 2 
\ddx 
{\cal L}_{\gamma \rho \pi \pi}} (x_1)
{\cal L}_{\rho \gamma} ( x_2)  \nonumber  \\
+{{i^3 \over 3!} 3 \dddx {\cal L}_{\rho\rho\pi\pi}}(x_1)
{\cal L}_{\rho\gamma}( x_2 ){\cal L}_{\rho\gamma}( x_3 )
] |\pi \rangle   \nonumber    \\
={e^2 g^2 \over 4}\langle \pi |i\dx (\pi_1^2(x)+\pi_2^2(x))
{1 \over g^2f_\pi^2}\{2F^2 g^{\mu\nu} {\Delta_F}^{(\gamma\rho)}
_{\mu\nu}(x-y)|_{x=y}   \nonumber \\
+{1 \over \pi^2}
g^{\mu\nu}g^{\lambda\rho}\p_\lambda^x\p_\rho^y 
{\Delta_F}^{(\gamma\rho)}_{\mu\nu}(x-y)|_{x=y}
\} |\pi \rangle ,
\end{eqnarray}
where
\begin{eqnarray}
{\Delta_{F}^{(\gamma\rho)}} _{\mu\nu}(x-y)&=&\dk
     {{\Delta_{F}} ^{(\gamma\rho)}} _{\mu\nu}
     (k)e^{-ik(x-y)}, \nonumber \\
{\Delta_F^{(\gamma\rho)}} _{\mu\nu}(k)&=&
 {-i \over k^2}[{m_\rho^4 \over (k^2-m_\rho^2 )^2}
 (g_{\mu\nu}-{k_\mu k_\nu \over k^2})+a{k_\mu k_\nu \over
  k^2}].
\end{eqnarray}
We call ${\Delta_F}^{(\gamma\rho)}_{\mu\nu}(x-y)$ as photon propagator
within $\rho$(to see Appendix A for details).

It is easy to check that the Eq.(22) can be re-obtain by
the following steps: at first, computing Fig.(1a) by using
${\cal L}_{\gamma\gamma\pi\pi}$, secondly, substituting
 ${\Delta_F}^{(\gamma\rho)}
_{\mu\nu}(x-y)$ for 
 ${\Delta_F}^{(\gamma)}
_{\mu\nu}(x-y)$ in it, then one reaches (22) again.
It is constructive that the substitution of
${\Delta_F}^{(\gamma)}_{\mu\nu} \longrightarrow
{\Delta_F}^{(\gamma\rho)}_{\mu\nu}$ in above is the
consequence of VMD. This rule is generally valid for
all VMD-process in the two-flavor case and it is useful for practical
calculations.

Using (16) and substituting (23) into (22), we get
the total contributions of Fig (1a), (1b) and (1c) to
$(m_{\pi^+}^2-m_{\pi^0}^2)$,
\begin{eqnarray}
(m_{\pi^+}^2-m_{\pi^0}^2)_1
={2i S_\pi (1) \over \langle \pi |\dx (\pi_1^2 +\pi_2^2 )
	     |\pi \rangle}  \nonumber  \\
=i{e^2 \over f_\pi^2} \dk (F^2+{k^2 \over 2\pi^2})
 [{m_\rho^4 \over k^2 (k^2-m_\rho^2)^2}(D-1)
   +{a \over k^2}],
\end{eqnarray}
where $D=4-\epsilon.$
According to the rule of dimensional regularization, i.e. 't Hooft-Veltman
Conjecture\cite{LH}, the last term
in (24) will vanish. Therefore $(m_{\pi^+}^2-m_{\pi^0}^2)_1 $
is gauge-independent.

Secondly, from Eqs.(19),(12),(15) and using substitution (9),
we have
\begin{eqnarray}
\lefteqn{
S_\pi (2)= {\langle \pi | T} \{ { i^2 \over  2!}
\int d^4x_1 d^4x_2 {\cal L}_{\pi  a
\gamma} ( x_{ 1}) 
{\cal L}_{\pi  a \gamma}( x_{ 2}) } \nonumber \\
& &+{i^3 \over 3!} 6 \int d^4x_1 d^4x_2 d^4x_3  
 {\cal L}_{\pi  a\rho}( x_1) 
{\cal L}_{\pi a\gamma}(x_2 )
{\cal L}_{\rho\gamma}(x_3)  \nonumber  \\
& & +{i^4 \over 4!} 6 \dddx d^4 x_4 {\cal L}_{\pi a\rho}(x_1)
{\cal L}_{\pi a \rho}(x_2){\cal L}_{\rho\gamma}(x_3)
{\cal L}_{\rho\gamma}(x_4) \} | \pi \rangle .
\end{eqnarray}
The straightforward calculation shows
\begin{equation}
S_\pi (2)=-{e^2 \gamma^2 \over 2g^2f_\pi^2}\langle \pi
|\int d^4p\pi_a(p) \pi_a(-p) (2\pi )^4\Gamma_2 (p^2)|\pi
\rangle
\end{equation}
where $a=1,2,$ and
\begin{eqnarray}
\pi_a(p)&=&{1 \over (2\pi)^4}\dx \pi_a(x)e^{-ipx}, \\
\Gamma_2(p^2)&=&\dk (F^2+{k^2 \over 2\pi^2})^2
{g^{\mu\nu}-{k^\mu k^\nu \over m_a^2} \over -k^2+m_a^2}
[{m_\rho^4 \over q^2(q^2-m_\rho^2)^2}
(g_{\mu\nu}-{q_\mu q_\nu \over q^2}) \nonumber \\
&&+a{q_\mu q_\nu \over q^4}],
\end{eqnarray}
here $q=p-k$. On $\pi-$mass shell, $p^2=m_\pi^2=0$
(chiral limit), so we have
\begin{equation}
S_\pi (2)=-{e^2 \gamma^2 \over 2g^2f_\pi^2}\langle \pi
|\dx\pi_a(x) \pi_a(x) |\pi
\rangle \Gamma_2(p^2=0)
\end{equation}
where $
\int d^4p\pi_a(p) \pi_a(-p) (2\pi )^4 =\dx \pi_a(x)
\pi_a(x) $ (to see
Eq.(27)) has been used.
From Eq.(28) the gauge-
dependent term of $\Gamma_2(p^2=0)$ is 
$$
a\dk (F^2+{k^2 \over 2\pi^2})^2
{1 \over m_a^2 k^2}
$$
This term equals to zero according to 't Hooft-Veltman Conjecture
in dimensional regularization. Therefore $S_\pi(2)$ is
gauge-independent. Thus, using (16), we get
\begin{eqnarray}
\lefteqn{
 (m_{\pi^+}^2-m_{\pi^0}^2)_2= -{ie^2\gamma^2 \over
g^2 f_\pi^2 } \Gamma_2(p^2=0)}  \nonumber \\
&& =i{{e^2 \gamma^2} \over {g^2 f_\pi^2}}\int\frac{d^4 k}{(2\pi)^4}
(F^2+{k^2 \over 2\pi^2} )^2
{{m_\rho^4 (D-1)} \over {k^2 (k^2-m_\rho^2 )^2
(k^2 -m_a^2)}}.
\end{eqnarray}

The $S_\pi (3) $ corresponding to Fig.(3) reads
\begin{eqnarray}
\lefteqn{
S_\pi (3)= {\langle \pi | T} \{ { i^2 \over  2!}
 \ddx {\cal L}_{\pi \pi
\gamma} ({\it x}_{\rm 1}) 
{\cal L}_{\pi \pi \gamma}({\it x}_{\rm 2}) } \nonumber \\
& &+{{i^3 \over 3!} 6 \dddx  
 {\cal L}_{\pi \pi \rho}}( x_1) 
{\cal L}_{\pi \pi \gamma}(x_2 )
{\cal L}_{\rho\gamma}(x_3)  \nonumber  \\
& & +{i^4 \over 4!} 6 \dddx d^4 x_4 {\cal L}_{\pi \pi \rho}(x_1)
{\cal L}_{\pi \pi \rho}(x_2){\cal L}_{\rho\gamma}(x_3)
{\cal L}_{\rho\gamma}(x_4) \} | \pi \rangle \nonumber \\
&=&-{e^2 \over 2F^4} \langle \pi |\dx \pi_a (x) \pi_a(x) |
\pi \rangle \Gamma_3 (p^2=0)
\end{eqnarray}
where
\begin{eqnarray}
\Gamma_3(p^2=0)&=&\dk (F^2+{k^2 \over 2\pi^2})^2
{k^\mu k^\nu \over k^2}
[{m_\rho^4 \over k^2(k^2-m_\rho^2)^2}
(g_{\mu\nu}-{k_\mu k_\nu \over k^2})
+a{k_\mu k_\nu \over k^4}]  \nonumber \\
&=&{ \dk   a } {1 \over k^2}  (F^2+ { k^2  \over { 2 \pi^2  }})^2
\end{eqnarray}
Using dimensional regularization, we have $\Gamma_3(p^2=0)=0$
, then $S_\pi(3)=0$ and
\begin{equation}
(m_{\pi^+}^2-m_{\pi^0}^2)_3=0.
\end{equation}

The total $\pi^+-\pi^0$ mass difference is the sum of
Eqs. (21), (30) and (33), which is
\begin{equation}
m_{\pi^+}^2-m_{\pi^0}^2 =i{e^2 \over f_\pi^2}
{\dk (D-1)m_\rho^4} {{(F^2+{k^2 \over 2\pi^2})} \over
{k^2 (k^2-m_\rho^2)^2 }}
[1+{\gamma^2 \over g^2}{{F^2+{k^2 \over 2\pi^2}} \over
{k^2 -m_a^2} }]
\end{equation}
The integration calculation for (34) is standard. We get the result of
\begin{eqnarray}
(m_{\pi^+}^2 -m_{\pi^0}^2)&=&{3\alpha_{
em} m_\rho^4 \over 8\pi f_\pi^2 }
\{{\gamma^2 \over g^2\pi^2 m_\rho^2} (F^2+ {m_a^2 \over 2\pi^2})
 \nonumber \\
&&+(2+ {\gamma^2 \over g^2 \pi^2})({F^2 \over m_\rho^2}
+{1 \over 3 \pi^2} -8 \chi_\rho )\nonumber\\
& &-{2\gamma^2 \over g^2 (m_a^2-m_\rho^2 )}
(F^2+{m_a^2 \over 2\pi^2})^2
({1 \over m_\rho^2} +{1 \over m_a^2-m_\rho^2} log {m_\rho^2 \over 
m_a^2}) \}
\end{eqnarray}
where $\alpha_{em}=e^2/4\pi ={1 \over 137}$ and
\begin{equation}
\chi_\rho= ({\mu^2 \over m_\rho^2 })^{\epsilon /2}
{1 \over (4\pi )^{D/2}} \Gamma (2-{D \over 2}).
\end{equation} 
It is essential that the logarithmic divergence in (35) (or (36)) can be
factorized by using the intrinsic parameter $g$ in this theory.
Comparing Eq.(3) with Eq.(36), we have
\begin{equation}
\chi_\rho=\frac{1}{8}g^2+{1 \over 32\pi^2}+{1 \over 16\pi^2}
log{f_\pi^2 \over 6(g^2 m_\rho^2-f_\pi^2)}.
\end{equation}
where Eq.(6) has been used. When $g$ is determined, $\chi_\rho$ will be fixed, and the final result
of (35) is finite.

The determination of $g$ can be done by taking $f_\pi$, $m_\rho$ and
$m_a$ as inputs.Substituting $f_\pi=0.186 GeV, m_\rho=
0.768 GeV $ and $ m_a=1.20GeV$ into Eqs.(4)and (6), we obtain
\begin{equation}
{g=0.39}
\end{equation}
Then
\begin{equation}
{m^2_{\pi^+}-m^2_{\pi^0}=0.001465GeV^2=2 m_\pi\times5.3 MeV}
\end{equation}
which is in reasonable agreement with the experimental value
of $2 m_\pi\times4.6 MeV$\cite{PDG}.

\section{$a_1^+-a_1^0$ electromagnetic mass difference}

The interaction
Lagrangians contributing to
$a_1^+-a_1^0 $ electromagnetic mass difference read
\begin{eqnarray}
{\cal L}_{\rho\rho a a}&=&-{2\gamma^2 
\over g^2} \rho_\mu^j \rho_\nu^k
( a^{j\mu}a^{k\nu}-g^{\mu\nu}a^{j\lambda}a^k_\lambda )  
+{\gamma^2 \over \pi^2 g^4}\rho_\mu^j \rho^{k\mu}
(\delta_{jk} a_\nu^i a^{i\nu}-a_\nu^j a^{k\nu} )  \\
&{\cal L}_{\rho a a}&={2 \over g}(1-{\gamma^2 \over g^2\pi^2})
\epsilon_{ijk}a_\mu^ja_\nu^k \p^\nu \rho^{i\mu}
-{2 \over g}\epsilon_{ijk}\rho_\mu^i a_\nu^j
(\p^\mu a^{k\nu}-\gamma^2\p^\nu a^{k\mu})   \\
&{\cal L}_{\rho a\pi}&=
\frac{2}{g}\epsilon_{ijk}\rho_\mu^i[c_1 \pi_j a^{k\mu}
+c_2 (\p_\nu \pi_j \p^\mu a^{k\nu} -a^k_\nu \p^\mu \p^\nu
\pi_j)]  \nonumber \\
&& +\frac{2}{g}\epsilon_{ijk} (\p_\mu \rho_\nu^i-\p_\nu \rho^i_\mu )
[c_3 \p^\mu (a^{k\nu}\pi_j )+c_4 \p^\mu \pi_j a^{k\nu}]
\end{eqnarray}
where
\begin{eqnarray}
&c_1&={\gamma \over f_\pi g}[F^2 +({1 \over 2\pi^2}
-2cg) m_a^2], \\
&c_2&={\gamma \over 2 f_\pi \pi^2 g} (1-{2c \over g}), \\
&c_3&={3\gamma \over 2 f_\pi \pi^2 g}(1-{2c \over g})+{2\gamma c\over f_\pi},\\
&c_4&={2\gamma c\over f_\pi }.
\end{eqnarray}
 The corresponding photon-$a_1$ interaction Lagrangians
${\cal L}_{\gamma\gamma a a}, {\cal L}_{\gamma a a }$ and
${\cal L}_{\gamma a \pi}$ can be constructed by the
substitution (9) and Eqs.(40)-(42). It is similar to the
preceding section that these Lagrangians and ${\cal L}_{\rho\gamma}$
(Eq.(12)) provide the dynamics for the mass splitting of $a_1$ due
to electromagnetic interactions. The Feynman diagrams are shown in
Figures (4), (5) and (6). The corresponding $S$-matrices are
denoted as $S_a(1), S_a(2)$ and $S_a(3)$, and
\begin{equation}
S_a=S_a(1)+S_a(2)+S_a(3).
\end{equation}
We calculate $S_a(1), S_a(2)$ and $S_a(3)$ separately in the
following.

For Fig.(4), from Eqs.(40) (9) and (12), we have
\begin{eqnarray}
\lefteqn{
S_a (1) = \langle a |T [i
\int d^4 x_1 {\cal L}_{ \gamma \gamma  a a}
( x_1 )
+{ i^2 \over  2! } 2 
\int d^4x_1 d^4x_2 
{\cal L}_{\gamma \rho  a a} (x_1)
{\cal L}_{\rho \gamma} ( x_2) } \nonumber  \\
&&+{i^3 \over 3!} 3 
\int d^4x_1 d^4x_2 d^4x_3 {\cal L}_{\rho\rho  a a}(x_1)
{\cal L}_{\rho\gamma}( x_2 ){\cal L}_{\rho\gamma}( x_3 )
] |a \rangle
\end{eqnarray}
Using (15), we get
\begin{eqnarray}
& &(m_{a^+}^2-m_{a^0}^2)_1\nonumber\\
&&=ie^2\frac{\gamma^2\langle a| \int d^4 x
a^{\underline{i}\mu}a^{\underline{i}\nu}|a\rangle-\langle a| \int d^4 x
a^{\underline{i}\lambda}a_{\lambda}^{\underline{i}}|a\rangle
g^{\mu\nu}}{\langle a| \int d^4 x a_{\mu}^{\underline{i}}a^{\underline{i}\mu}
|a\rangle}
\nonumber \\
& &\int\frac{d^4 k}{(2\pi)^4}\frac{m_\rho^4}{k^2(k^2-m_\rho^2)^2}
(g_{\mu\nu}-\frac{k_\mu k_\nu}{k^2})
\end{eqnarray}
where $\underline{i}=1,2$.

For Fig.(5),from Eqs.(41),(9) and (12), we have
\begin{eqnarray}
\lefteqn{
S_a (2)= \langle a | T \{ { i^2 \over  2!}
\int d^4x_1 d^4x_2 {\cal L}_{a a
\gamma} ( x_1) 
{\cal L}_{ a a \gamma}( x_2) } \nonumber \\
& &+{i^3 \over 3!} 6 
\int d^4x_1 d^4x_2 d^4x_3  
 {\cal L}_{  a a\rho}( x_1) 
{\cal L}_{a a\gamma}(x_2 )
{\cal L}_{\rho\gamma}(x_3)  \nonumber  \\
& & +{i^4 \over 4!} 6 \dddx d^4 x_4 {\cal L}_{a a\rho}(x_1)
{\cal L}_{a a \rho}(x_2){\cal L}_{\rho\gamma}(x_3)
{\cal L}_{\rho\gamma}(x_4) \} | a \rangle .
\end{eqnarray}
Using Eq.(15), we obtain
\begin{eqnarray}
& &(m_{a^+}^2-m_{a^0}^2)_2\nonumber \\
& &=\frac{ie^2}{\langle a| \int d^4 x a^{\underline{i}\mu}
a_{\mu}^{\underline{i}}|a\rangle}\int\frac{d^4 k}{(2\pi)^4}\frac{1}{k^2-2p\cdot k}
\frac{m_\rho^4}{k^2(k^2-m_\rho^2)^2}\nonumber \\
& &\{\langle a| \int d^4 x a^{\underline{i}\mu}a_{\mu}^{\underline{i}}|a\rangle
[4 m_a^2+(b^2+2b\gamma^2)k^2
+2\gamma^4 p\cdot k-\frac{4(p\cdot k)^2}{k^2}\nonumber \\
& &-\frac{1}{m_a^2}(b k^2-(b-\gamma^2)p\cdot k)^2]
+\langle a| \int d^4 x a_{\mu}^{\underline{i}}a_{\nu}^{\underline{i}}|a\rangle
k^\mu k^\nu [-(3b^2-4 b+4)\nonumber \\
& &+D(b+\gamma^2)^2+4\gamma^2
-6b\gamma^2-2\gamma^4-\frac{2\gamma^4 p\cdot k}{k^2}
+\frac{1}{m_a^2 k^2}(b k^2-2(1-\gamma^2)p\cdot k)^2]\}
\end{eqnarray}
where $b=1-\frac{\gamma^2}{\pi^2 g^2}$, and
$p$ is the external momentum of $a_1$-fields.
The Fourier transformation for mass-shell $a_1$-fields
is
$$
a_\mu^i(p)={1\over (2\pi)^4}\int d^4x a_\mu^i(x)
e^{-ipx}
$$
with
\begin{eqnarray}
p^2=m_a^2,\;\;\;\;\;\;{\rm and }\;\;p^\mu a_\mu^i(p)=0.
\end{eqnarray}

For Fig.(6), from Eqs. (42), (9) and (12). we have
\begin{eqnarray}
\lefteqn{
S_a (3)= \langle a | T \{ { i^2 \over  2!}
\int d^4x_1 d^4x_2 {\cal L}_{a \pi
\gamma} ( x_1) 
{\cal L}_{ a \pi \gamma}( x_2) } \nonumber \\
& &+{i^3 \over 3!} 6 
\int d^4x_1 d^4x_2 d^4x_3  
 {\cal L}_{  a \pi \rho}( x_1) 
{\cal L}_{a \pi\gamma}(x_2 )
{\cal L}_{\rho\gamma}(x_3)  \nonumber  \\
& & +{i^4 \over 4!} 6 \dddx d^4 x_4 {\cal L}_{a \pi\rho}(x_1)
{\cal L}_{a \pi \rho}(x_2){\cal L}_{\rho\gamma}(x_3)
{\cal L}_{\rho\gamma}(x_4) \} | a \rangle .
\end{eqnarray}
and
\begin{eqnarray}
& &(m_{a^+}^2-m_{a^0}^2)_3\nonumber \\
& &=\frac{-ie^2}{\langle a| \int d^4 x a_{\mu}^{\underline{i}}
a^{\underline{i}\mu}|a\rangle}\int\frac{d^4 k}{(2\pi)^4}\frac{1}{(p-k)^2}
\frac{m_\rho^4}{k^2(k^2-m_\rho^2)^2}\nonumber\\
& &\{\langle a| \int d^4 x a_{\mu}^{\underline{i}}a^{\underline{i}\mu}|a\rangle
(c_1-3 c_2 p\cdot k+
c_3 k^2)^2+\nonumber \\
& &\langle a| \int d^4 x a_{\mu}^{\underline{i}}a_{\nu}^{\underline{i}}|a\rangle k^\mu k^\nu
[c_2 m_{a}^2-\frac{(c_1-2 c_2 p\cdot k+c_3 k^2)^2}{k^2}
]\}
\end{eqnarray}

It needs to be checked that $(m_{a^+}^2-m_{a^0}^2)_{1,2,3}$
are gauge-independent. The gauge-dependent terms
of $(m_{a^+}^2-m_{a^0}^2)_1$, which come from Fig.4a,
will vanish according to the rule of dimensional regularization.

The gauge dependent terms in $S_a(2)$(to be denoted as
$S_a(2)_G$) come from Fig.5a. Using VMD,the correspondent photon-meson
interaction Lagrangians is
\begin{equation}
{\cal L}_{\gamma a a}=e b\epsilon_{3jk}a_{\mu}^j a_{\nu}^k
\partial^{\nu} A^\mu-e\epsilon_{3jk}A_{\mu} a_{\nu}^j(
\partial^{\mu} a^{k\nu}-\gamma^2\partial^{\nu} a^{k\mu})
\end{equation}
Then
\begin{eqnarray}
S_a(2)_G&=& a^\prime \frac{e^2}{2}\langle a| \int d^4 x a_{\mu}^{\underline{i}}
 a^{\underline{i}\mu}|a\rangle
\int \frac{d^4 k}{(2\pi)^4}\frac{1}{k^2}\{1-\frac{2p\cdot k}{k^2}\nonumber \\
&&+\frac{k^2}{D}[-\frac{\gamma^4}{k^2}-\frac{(1-\gamma^2)^2}{m_a^2 k^2}(
k^2-2 p\cdot k)]\}
\end{eqnarray}
where $a^\prime$ is gauge parameter. 't Hooft-Veltman Conjecture  will make sure that
$S_a(2)$ is gauge independent.

The photon-meson interaction Lagrangian contributing to the gauge dependent
term $S_a(3)_G$(Fig.6a) is
\begin{eqnarray}
{\cal L}_{\gamma a \pi}&=&e\epsilon_{3jk}A_\mu[c_1 \pi_j a^{k\mu}
+c_2(\partial^\nu\pi_j\partial^\mu a_{\nu}^k-a^{k\nu}
\partial^\mu\partial_\nu\pi_j)]
\end{eqnarray}
We will have
\begin{eqnarray}
S_a(3)_G&=&-a^\prime \frac{e^2}{2}\langle a|
 \int d^4 x a^{\underline{i}}_{\mu}a^{\underline{i}\mu}|a\rangle
\int \frac{d^4 k}{(2\pi)^4}\frac{1}{k^2}[\frac{(c_1-c_2 m_a^2)^2}{k^2(p-k)^2}
\nonumber \\
&&+\frac{c_2(c_1-c_2 m^2_a)}{k^2}+\frac{c_2^2(p-k)^2}{k^2}]
\end{eqnarray}
The third term will vanish because of dimensional regularization.
By Eq.(5) and the definition of $c_1$ and $c_2$, we obtain that
$c_1-c_2 m_a^2=0$. Thus the $S_a(3)_G=0$.

The $g$ has been determined in Eq.(38) and the logarithmic divergences
in the above Feynman integrations(Eqs.(49),(51) and (54))
can also be factorized by using Eq.(37),
so there are no any further
unknown parameters in the expressions of
$(m_{a^+}^2-m_{a^0}^2)_{1,2,3}$.
After a long but straightforward calculation,
we can get the final results for $(m_{a^+}^2-m_{a^0}^2)_{1,2,3}$,
whose form is very tedious.
The numerical results for them are
\begin{eqnarray}
(m_{a^+}^2-m_{a^0}^2)_1&=&-0.000648GeV^2,\nonumber  \\
(m_{a^+}^2-m_{a^0}^2)_2&=&-0.002688GeV^2, \nonumber \\
(m_{a^+}^2-m_{a^0}^2)_3&=&0.001896GeV^2.
\end{eqnarray}
Totally,
\begin{equation}
(m_{a^+}^2-m_{a^0}^2)_{EM}=-0.001440GeV^2=-2 m_a\times0.57MeV
\end{equation}

\section{$K^+$-$K^0$ electromagnetic mass difference and the violation of
Dashen's theorem}

In this Section and next two Sections, our method is extended to the studies
of the electromagnetic self-energies of the strange-flavor mesons.
As mentioned above, the large strange quark
mass will result in the SU(3) symmetry breaking playing an important
role in these calculations. Dashen's theorem, which states that
the electromagnetic contributions to the difference between the mass square
of kaons and pions are equal, is valid only in the chiral SU(3) limit.
Corrections to the electromagnetic self-energies to the leading order
in quark mass expansion are sure to lead to the violation of
Dashen's theorem.
Therefore, it is necessary to evaluate the electromagnetic self-energies
of the strange-flavor mesons and the corrections to Dashen's theorem
to the order of $m_s$ or $m_K^2$.

From Eq.(3) in Ref.\cite{Li2}(${\cal L}_{RE}$),
the interaction Lagrangians which can contribute to
electromagnetic mass difference between $K^+$ and $K^0$ are
\begin{eqnarray}
{\cal L}_{KKvv}&=&\frac{1}{f_k^2 g^2}\{2 F^2 \rho^{3\mu}
v_\mu^8(K^{+} K^{-}-K^0\bar{K}^0)
\nonumber \\
& &+\frac{1}{\pi^2}\partial_\nu
\rho^{3\mu}\partial^\nu v_\mu^8(K^+K^{-}-K^0\bar{K}^0)
\nonumber \\
& &+[\frac{1+(1-\frac{2 c^{\prime}}{g})^2}
{\pi^2}-8 c^{\prime 2}]\rho^{3\mu} v_\mu^8(\partial_\nu
K^+ \partial^\nu K^--\partial_\nu K^0
\partial^\nu \bar{K}^0)\nonumber \\
& &-\frac{2(1-\frac{2c^\prime}{g})}{\pi^2}\rho_{\nu}^3 v_\mu^8
(K^+\partial^{\mu\nu}K^{-}-
K^0\partial^{\mu\nu}\bar{K}^0+ h.c.)\nonumber \\
& &+4 c^{\prime 2}\rho^3_\mu v_\nu^8
(\partial^{\mu} K^+\partial^{\nu} K^{-}
-\partial^\mu K^0
\partial^\nu\bar{K}^0 + h.c.)\}
\end{eqnarray}
\begin{eqnarray}
{\cal L}_{KKv}&=&\frac{i}{g}\alpha_1[\rho_\mu^3(K^+\partial^\mu K^{-}-
K^0\partial^\mu\bar{K}^0)+
v_\mu^8(K^+\partial^\mu K^{-}+K^0\partial^\mu\bar{K}^0)]
\nonumber \\
& &-\frac{i}{g}\alpha_2[\rho^3_\mu(K^+\partial^2\partial^\mu K^{-}-K^0\partial^2
\partial^\mu \bar{K}^0)+
v_\mu^8(K^+\partial^2\partial^\mu K^{-}+K^0\partial^2
\partial^\mu \bar{K}^0)]
\nonumber \\
& &+\frac{i}{g}\alpha_3[\rho^3_\mu(\partial^{\mu\nu} K^+\partial_\nu K^--
\partial^{\mu\nu}K^0\partial_\nu\bar{K}^0)
+v_\mu^8(\partial^{\mu^\nu} K^+\partial_\nu K^-+
\partial^{\mu\nu}K^0\partial_\nu\bar{K}^0)]\nonumber \\
& &+h.c.
\end{eqnarray}
\begin{eqnarray}
{\cal L}_{K K_1 v}&=&\frac{i}{g}\beta_1[\rho_\mu^3(K^{+}K_1^{-\mu}-
K^0\bar{K}_1^{0\mu})+v_\mu^8(K^{+}K_1^{-\mu}+
K^0\bar{K}_1^{0\mu}]\nonumber \\
& &+\frac{i}{g}\beta_2[\rho_\nu^3(\partial^{\mu\nu} K^+K_{1\mu}-
\partial^{\mu\nu} K^0\bar{K}^0_{1\mu})+v_\nu^8
(\partial^{\mu\nu} K^{+}K_{1\mu}+\partial^{\mu\nu} K^0
\bar{K}^0_{1\mu})]\nonumber \\
& &+\frac{i}{g}\beta_3[\rho_\mu^3(K_1^{+\mu}\partial^2 K^{-}-
K_1^{0\mu}\partial^2 \bar{K}^0)+v_\mu^8
(K_1^{+\mu}\partial^2 K^{-}+K_1^{0\mu}\partial^2\bar{K}^0)]\nonumber \\
& &-\frac{i}{g}\beta_4[\rho_\mu^3(K^{+}\partial^2 K_1^{-\mu}-
K^{0}\partial^2 \bar{K}_1^{0\mu})+v_\mu^8
(K^{+}\partial^2 K_1^{-\mu}+K^{0}\partial^2\bar{K}_1^{0\mu})]\nonumber \\
& &-\frac{i}{g}\beta_3[\rho_\nu^3(K^+\partial^{\mu\nu} K_{1\mu}^-
-K^0\partial^{\mu\nu}\bar{K}_{1\mu}^0)+v_\nu^8
(K^+\partial^{\mu\nu} K_{1\mu}^-+
K^0\partial^{\mu\nu}\bar{K}_{1\mu}^0)]\nonumber \\
& &-\frac{i}{g}\beta_5[\rho_\mu^3(\partial_\nu K_{1\mu}^+ \partial^\nu K^-
-\partial_\nu K_{1\mu}^0 \partial^\nu \bar{K}^0)+
v_\mu^8(\partial_\nu K_{1\mu}^+\partial^\nu K^-
+\partial_\nu K_{1\mu}^0 \partial^\nu \bar{K}^0)]\nonumber \\
& &+ h.c.
\end{eqnarray}
with
\begin{eqnarray}
\alpha_1&=&\frac{(1-\frac{2 c^{\prime}}{g})F^2}{f_k^2}, \nonumber \\
\alpha_2&=&\frac{(1-\frac{2 c^{\prime}}{g})}{2\pi^2 f_k^2}+
\frac{3(1-\frac{2c^{\prime}}{g})^2}{2\pi^2 f_k^2}-\frac{4{c^\prime}^2}{f_k^2},
\nonumber \\
\alpha_3&=&{(1-\frac{2c^{\prime}}{g})^2\over \pi^2 f_k^2}-
{4{c^\prime}^2\over f_k^2}.
\end{eqnarray}
and
\begin{eqnarray}
\beta_1={\gamma F^2\over g f_k}, \;\;\;\; \beta_2={\gamma\over
2\pi^2 g f_k} (1-\frac{2c^{\prime}}{g}),\nonumber\\
\beta_3={3\gamma\over 2\pi^2 g f_k}(1-\frac{2c^{\prime}}{g})+
{2\gamma c^{\prime}\over f_k},\;\;\;\beta_4={\gamma\over 2\pi^2 g f_k},
\nonumber \\
\beta_5={3\gamma\over 2\pi^2 g f_k}(1-\frac{2c^{\prime}}{g})+
{4\gamma c^{\prime}\over f_k}.
\end{eqnarray}
Where $v$ denotes the vector mesons including $\rho$,$\omega$ and $\phi$.
$v_\mu^8=\omega_\mu-\sqrt{2}\phi_\mu$,
$\partial^{\mu\nu}=\partial^\mu\partial^\nu$. Distinguishing from the
case of massless pions system, the nonzero strange
quark mass, i.e. $m_k^2\not=0$, will bring about the
contributions to $m^2_{K^+}-m^2_{K^0}$ from the abnormal
part of the effective Lagrangian. These vertices have been
found by the evaluation of $\frac{1}{g}K^*_{a\mu}\langle\bar{\psi}
\lambda_a\gamma^\mu \psi\rangle$ in Ref.\cite{Li2}.
\begin{eqnarray}
{\cal L}_{K^* K v}&=&-{3\over 2\pi^2 g^2}\frac{2}{f_k}
\epsilon^{\mu\nu\alpha\beta}K^+_{\mu}\partial_\beta K^-(\frac{1}{2}
\partial_\nu \rho_\alpha^3+\frac{1}{2}\partial_\nu\omega_\alpha
+\frac{\sqrt{2}}{2}\partial_\nu\phi_\alpha)\nonumber \\
& &-{3\over 2\pi^2 g^2}\frac{2}{f_k}\epsilon^{\mu\nu\alpha\beta}
K^0_\mu\partial_\beta\bar{K}^0(-\frac{1}{2}\partial_\nu\rho^3_\alpha+
\frac{1}{2}\partial_\nu\omega_\alpha
+\frac{\sqrt{2}}{2}\partial_\nu\phi_\alpha)\nonumber \\
& &+ h.c.
\end{eqnarray}
Here, we adopt the following definitions for the strange-flavor mesons.
\begin{eqnarray}
K^{\pm}=\frac{1}{\sqrt{2}}(K^4\pm i K^5),\;\;\;\;
K^0(\bar{K}^0)=\frac{1}{\sqrt{2}}(K^6\pm i K^7),\nonumber \\
K_{1\mu}^{\pm}=\frac{1}{\sqrt{2}}(K_{1\mu}^4\pm i K_{1\mu}^5),\;\;\;
K_{1\mu}^0(\bar{K}_{1\mu}^0)=\frac{1}{\sqrt{2}}(K_{1\mu}^6\pm i K_{1\mu}^7),
\nonumber \\
K_{\mu}^{\pm}=\frac{1}{\sqrt{2}}(K_\mu^4\pm i K_\mu^5),\;\;\;
K_{\mu}^0(\bar{K}_\mu^0)=\frac{1}{\sqrt{2}}(K_\mu^6\pm i K_\mu^7).
\end{eqnarray}
The interaction Lagrangians between the photon and $K$-meson
${\cal L}_{\gamma\gamma K K},{\cal L}_{\gamma K K},{\cal L}_{\gamma K_1 K}$,
and ${\cal L}_{\gamma K^* K}$  can be
obtained by the substitutions (9)-(11) and Eqs.(61)-(63) and (66).
The Feynman diagrams contributing to the electromagnetic mass difference
between $K^+$ and $K^0$ are shown in Figs.(7),(8),(9),(10). The corresponding
$S$-matrices are denoted as $S_K(1)$,$S_K(2)$,$S_K(3)$ and $S_K(4)$
respectively.

In Sec.2, we obtain $S_\pi$ by substituting
$\Delta_{F_{\mu\nu}}^{(\gamma\rho)}(x-y)$ for
$\Delta_{F_{\mu\nu}}^{(\gamma)}(x-y)$ after computing Fig.(1a) (2a) (3a).
Here, the involved vector mesons are not only $\rho$-mesons,but also $\omega$
and $\phi$-mesons. So it is not as simple as in the case of pions.
Practical calculations will show that we can get $S_K$ by changing the
form of this substitution(to see Appendix B).
Specifically, for $S_K(1),S_K(2),S_K(3)$ coming from the ${\cal L}_{RE}$, the
corresponding
propagator of the substitution should be $\Delta_{F_{1\mu\nu}}^{(\gamma v)}$
instead of $\Delta_{F_{\mu\nu}}^{(\gamma\rho)}$
\begin{eqnarray}
\Delta_{F_{1\mu\nu}}^{(\gamma v)}(x-y)&=&\int\frac{d^4 k}{(2\pi)^4}
\Delta_{F_{1\mu\nu}}^{(\gamma v)}(k) e^{-i k(x-y)},\nonumber \\
\Delta_{F_{1\mu\nu}}^{(\gamma v)}(k)&=&\frac{-i}{k^2}\{
[\frac{1}{3}\frac{m_\rho^2 m_\omega^2}{(k^2-m_\rho^2)(k^2-m_\omega^2)}+
 \frac{2}{3}\frac{m_\rho^2 m_\phi^2}{(k^2-m_\rho^2)(k^2-m_\phi^2)}]
 \nonumber \\
& &\times(g_{\mu\nu}-\frac{k_\mu k_\nu}{k^2})+a\frac{k_\mu k_\nu}{k^2}\}
\end{eqnarray}
Obviously, under the SU(3) limit, $m_\rho=m_\omega=m_\phi$,
$\Delta_{F_{1\mu\nu}}^{(\gamma v)}$ will go back
$\Delta_{F_{\mu\nu}}^{(\gamma \rho)}$.
However, for $S_K(4)$, which receives the contributions from the abnormal
part  Lagrangian ${\cal L}_{IM}$, the substituting propagator should be
$\Delta_{F_{2\mu\nu}}^{(\gamma v)}$,
\begin{eqnarray}
\Delta_{F_{2\mu\nu}}^{(\gamma v)}(x-y)&=&\int\frac{d^4 k}{(2\pi)^4}
\Delta_{F_{2\mu\nu}}^{(\gamma v)}(k) e^{-i k(x-y)},\nonumber \\
\Delta_{F_{2\mu\nu}}^{(\gamma v)}(k)&=&\frac{-i}{k^2}\{
[\frac{1}{3}\frac{m_\rho^2 m_\omega^2}{(k^2-m_\rho^2)(k^2-m_\omega^2)}-
 \frac{2}{3}\frac{m_\rho^2 m_\phi^2}{(k^2-m_\rho^2)(k^2-m_\phi^2)}]
 \nonumber \\
& &\times(g_{\mu\nu}-\frac{k_\mu k_\nu}{k^2})+a\frac{k_\mu k_\nu}{k^2}\}
\end{eqnarray}
Note that $\Delta_{F_{2\mu\nu}}^{(\gamma v)}$ is different from
$\Delta_{F_{1\mu\nu}}^{(\gamma v)}$(to see Appendix B).

Thus, it is easy to obtain the contributions of Figs.(7),(8),(9),(10)
to $(m_{K^+}^2-m_{K^0}^2)$ respectively.

Contribution of Fig.(7) is
\begin{eqnarray}
(\Delta m_K^2)_1&=&(m_{K^+}^2-m_{K^0}^2)_1=\frac{iS_K(1)}
{\langle K| \int d^4 x K^+ K^-|K\rangle}\nonumber\\
&=&i\frac{e^2}{f_k^2}\int \frac{d^4 k}{(2\pi)^4}(F_K^2+\frac{k^2}{2\pi^2})
(D-1)[\frac{1}{3}\frac{m_\rho^2 m_\omega^2}{k^2(k^2-m_\rho^2)(k^2-
m_\omega^2)}\nonumber \\
& &+\frac{2}{3}\frac{m_\rho^2 m_\phi^2}{k^2(k^2-m_\rho^2)(k^2-
m_\phi^2)}]
\end{eqnarray}
with
\begin{eqnarray*}
F_K^2=F^2+[\frac{1+(1-\frac{2c^\prime}{g})+(1-\frac{2c^\prime}{g})^2}{2\pi^2}
-3 {c^\prime}^2] p^2
\end{eqnarray*}
where $p$ is the external momentum of kaons, and $p^2=m_K^2$ on $K$-mass shell.

Contribution of Fig.(8) is
\begin{eqnarray}
(\Delta m_K^2)_2&=&(m_{K^+}^2-m_{K^0}^2)_2=\frac{iS_K(2)}{
\langle K| \int d^4 x K^+ K^-|K\rangle}\nonumber\\
&=&-ie^2\int\frac{d^4 k}{(2\pi)^4}
\frac{X^\mu X^\nu(g_{\mu\nu}-\frac{k_\mu k_\nu}{k^2})}{(p-k)^2-m_K^2}\nonumber\\
& &\times[\frac{1}{3}\frac{m_\rho^2 m_\omega}{(k^2-m_\rho^2)(k^2-m_\omega^2)}
+\frac{2}{3}\frac{m_\rho^2 m_\phi^2}{(k^2-m_\rho^2)(k^2-m_\phi^2)}]
\end{eqnarray}
with
\begin{eqnarray}
& &X_\mu=\alpha_1(q_\mu+p_\mu)+\alpha_2(q^2 q_\mu+p^2 p_\mu)-\alpha_3(p\cdot q)
(q_\mu+p_\mu),\\
& &q=p-k.\nonumber
\end{eqnarray}
	      
Contribution of Fig.(9) is
\begin{eqnarray}
(\Delta m_K^2)_3&=&(m_{K^+}^2-m_{K^0}^2)_3=\frac{iS_K(3)}{
\langle K| \int d^4 x K^+ K^-|K\rangle} \nonumber \\
&=&ie^2\int\frac{d^4 k}{(2\pi)^4} Y_{\mu\nu}\frac{g^{\mu\nu}-
\frac{q^\mu q^\nu}{m_{K_1}^2}}{q^2-m_{K_1}^2}\nonumber\\
& &\times[\frac{1}{3}\frac{m_\rho^2 m_\omega^2}{k^2(k^2-m_\rho^2)(k^2-m_\omega^2)}
+\frac{2}{3}\frac{m_\rho^2 m_\phi^2}{k^2(k^2-m_\rho^2)(k^2-\phi^2)}]
\end{eqnarray}
where
\begin{eqnarray}
Y_{\mu\nu}&=&(\beta_1+\beta_3 p^2+\beta_4 q^2-\beta_5 p\cdot q)^2(g_{\mu\nu}-
\frac{k_\mu k_\nu}{k^2})-2(\beta_1+\beta_4 q^2)\beta_2 p_\mu p_\nu\nonumber \\
& &+\beta_3^2 q_\mu q_\nu(p^2-\frac{(p\cdot k)^2}{k^2})+2(\beta_1+\beta_4
q^2-\beta_5 p\cdot q)\beta_3 p_\mu q_\nu \nonumber \\
& &-2(\beta_1+\beta_4 q^2-\beta_5 p\cdot q)(\beta_2 \frac{p\cdot k}{k^2}
p_\nu q_\mu+\beta_3 \frac{p\cdot k}{k^2} k_\mu q_\nu)
\end{eqnarray}
and $q=p-k$.

Contribution of Fig.(10) is
\begin{eqnarray}
(\Delta m_K^2)_4&=&(m_{K^+}^2-m_{K^0}^2)_4=\frac{iS_K(4)}{
\langle K| \int d^4 x K^+ K^-|K\rangle}\nonumber \\
&=&-\frac{9 i e^2}{2\pi^4 g^2 f_k^2}\int\frac{d^4 k}{(2\pi)^4}\frac{p^2 k^2
-(p\cdot k)^2}{(p-k)^2-m_{K^*}^2} \nonumber \\
& &\times[\frac{1}{3}\frac{m_\rho^2 m_\omega^2}{k^2(k^2-m_\rho^2)(k^2-m_\omega^2)}
-\frac{2}{3}\frac{m_\rho^2 m_\phi^2}{k^2(k^2-m_\rho^2)(k^2-m_\phi^2)}]
\end{eqnarray}

The gauge-independence of $(m_{K^+}^2-m_{K^0}^2)_{1,2,3,4}$ should be examined.
The gauge dependent terms in $(m_{K^+}^2-m_{K^0}^2)_1$ will
vanish according to 't Hooft-Veltman Conjecture,
which are similar to the cases of $S_\pi(1)$ and $S_a(1)$.

The gauge independent terms in $S_K(2)$(to be denotes as $S_K(2)_G$)
come from Fig.(8a). Using VMD, ${\cal L}_{\gamma K K}$ can be constructed
from ${\cal L}_{K K v}$. Thus we have
\begin{equation}
S_K(2)_G=-a e^2\langle K| \int d^4 x K^+ K^-|K\rangle\int \frac{d^4 k}{(2\pi)^4}\frac{
X_\mu X_\nu k^\mu k^\nu}{(q^2-m_K^2) (k^2)^2}
\end{equation}
From Eq.(72), we have
\begin{eqnarray*}
X_\mu k^\mu=-\alpha_1(q^2-p^2)-\alpha_2(p^2-p\cdot k+k^2)(q^2-p^2)+
\alpha_3(p^2-p\cdot k)(q^2-p^2)
\end{eqnarray*}
Mass shell condition leads to $p^2$=$m_K^2$,
so the term $(q^2-m_K^2)$ in the denominator of $S_K(2)_G$ will be
reduced. This  means that the contribution of $S_K(2)_G$ is zero
in the framework of the dimensional regularization.

Likewise, we will obtain $S_K(3)_G$(Fig.(9a)), which is
\begin{eqnarray}
S_K(3)_G&=&a e^2\langle K| \int d^4 x K^+ K^-|K\rangle\int\frac{d^4 k}{(2\pi)^4}\frac{1}{q^4}
\frac{g^{\mu\nu}-\frac{k^\mu k^\nu}{m_{K_1}^2}}{k^2-m_{K_1}^2}
\nonumber \\
& &\times (W_1 p_\mu-W_2 k_\mu)(W_1 p_\nu-W_2 k_\nu)\nonumber \\
&=& a e^2\langle K| \int d^4 x K^+ K^-|K\rangle\int\frac{d^4 k}{(2\pi)^4}\frac{1}{q^4}
\{W_1^2\frac{(p^2-\frac{(p\cdot k)^2}{m_{K_1}^2})}{k^2-m_{K_1}^2}+
\nonumber \\
& & 2 W_1 W_2 \frac{p\cdot k}{m_{K_1}^2}-W_2^2\frac{k^2}{m_{K_1}^2}\}
\end{eqnarray}
where
\begin{eqnarray}
& &W_1=\beta_1+(\beta_4-\beta_2-\beta_6)k^2+\beta_2 m_K^2, \nonumber \\
& &W_2=\beta_1+\beta_4 k^2-\beta_6 p\cdot k, \nonumber \\
& &\beta_6={2\gamma c^\prime}{f_k}. \nonumber
\end{eqnarray}
The contributions of the second and third terms in Eq.(77) are zero because of
't Hooft-Veltman Conjecture. Since our calculations are only to the order of
$m_K^2$, the denominator of the first term in Eq.(77), $k^2-m_{K_1}^2$ can also
be reduced. Here, a relation $\beta_1+(\beta_4-\beta_2-\beta_6)m_{K_1}^2=0$
,which can be easily obtained by Eq.(7), has been used. Thus, the $S_K(3)$
is gauge independent.

The gauge dependent terms $S_K(4)_G$(Fig.(10a)), which receive contributions
from the abnormal part of the effective Lagrangian
${\cal L}_{K K^* \gamma}$, are in the following
\begin{eqnarray}
S_K(4)_G&=&-a\frac{3 e^2}{4\pi^4 g^2 f_k^2}\langle K| \int d^4 x K^+ K^-|K\rangle\int
\frac{d^4 q}{(2\pi)^4}p_\beta p_{\beta^\prime} \epsilon^{\mu\nu\alpha\beta}
\epsilon^{\mu^{\prime}\nu^{\prime}\alpha^{\prime}\beta^{\prime}}\nonumber \\
& &\times \frac{g_{\mu\mu^\prime}-\frac{(p-q)_\mu (p-q)_{\mu^{\prime}}}
{m_{K^*}^2}}{(p-q)^2-m_{K^*}^2}\frac{q_\nu q_{\nu^\prime} q_{\alpha}
q_{\alpha^\prime}}{q^4}
\end{eqnarray}
It is obvious that $S_K(4)_G$ will vanish because of the totally
antisymmetric tensor $\epsilon_{\mu\nu\alpha\beta}$.

From Eqs.(70)(71)(73)(75), It is not difficult to conclude that the contributions of
$S_K(2)$ and $S_K(4)$ are proportional to $p^2$. So in the chiral limit,
$p^2=m_K^2=0$, only $S_K(1)$ and $S_K(3)$ contribute to $(m_{K^+}^2-
m_{K^0}^2)$.
Then, we have
\begin{eqnarray}
{\Delta m_K^2}_{m_s=0}&=&i\frac{e^2}{f_k^2}\int\frac{d^4 k}{(2\pi)^4}
(D-1)(F^2+\frac{k^2}{2\pi^2})(1+\frac{\gamma^2}{g^2}\frac{F^2
+\frac{k^2}{2\pi^2}}{k^2-m_{K_1}^2})\nonumber \\
& &\times[\frac{1}{3}\frac{m_\rho^2 m_\omega^2}
{k^2(k^2-m_\rho^2)(k^2-m_\omega^2)}+\frac{2}{3}\frac{m_\rho^2 m_\phi^2}
{k^2(k^2-m_\rho^2)(k^2-m_\phi^2)}]
\end{eqnarray}
Taking $f_k=f_\pi$, $m_\rho=m_\omega=m_\phi$, and $m_{K_1}=m_a$,
 the above equation reduces to Eq.(34).
This indicates that Dashen's theorem is automatically obeyed
in the chiral SU(3) limit of the present theory. However,
SU(3) symmetry breaking effects will lead to the violation of Dashen's
theorem. The total $\Delta m_K^2$(the sum of $(m_{K^+}^2-m_{K^0}^2)_{1,2,3,4}$)
which is evaluated to the order of $m_K^2$ can be read off from Eqs.(70),(71),
(73),(75).
It is straightforward to perform these Feynman integrations,
although the calculating processes and the results are not as simple as
that in the case of pions. We don't present the final expressions of $
(m_{K^+}^2-m_{K^0}^2)_{1,2,3,4}$ here.
Note that only the
logarithmic divergences are
involved in the above Feynman integrations,
which can be factorized by using Eq.(37). $f_k$ is determined from
Eqs.(7)(8), not as an input, and $g=0.39$ still holds.
Numerically, the results of Eqs.(70)(71)(73)(75) are
\begin{eqnarray*}
& &(m_{K^+}^2-m_{K^0}^2)_1=0.002193GeV^2,\\
& &(m_{K^+}^2-m_{K^0}^2)_2=-0.000430GeV^2,\\
& &(m_{K^+}^2-m_{K^0}^2)_3=0.000571GeV^2,\\
& &(m_{K^+}^2-m_{K^0}^2)_4=0.000139GeV^2.
\end{eqnarray*}
Totally, we have
\begin{equation}
({\Delta m_K^2})_{EM}=(m_{K^+}^2-m_{K^0}^2)_{EM}=0.002473GeV^2
=2 m_K \times 2.5 MeV
\end{equation}
Then the correction to Dashen's theorem beyond the chiral limit is
\begin{eqnarray}
& &\rho_{EM}={(m_{K^+}^2-m_{K^0}^2)_{EM}\over(m_{\pi^+}^2-m_{\pi^0}^2)_{EM}}
=1.68,\nonumber \\
& &(\Delta m_K^2)_{EM}-(\Delta m_\pi^2)_{EM}=1.08\times 10^{-3} GeV^2.
\end{eqnarray}
The results show rather large violation of Dashen's theorem, which
is in correspondence with the one by Donoghue et al.\cite{DHW} and
Bijnens et al.\cite{BB,JB}.

It has been known that mass difference between $K^+$
and $K^0$ receives the contributions from both electromagnetic self-energy
and mass difference of $m_u$ and $m_d$, i.e.
\begin{equation}
(m_{K^+}^2-m_{K^0}^2)_{EXPT}=(m_{K^+}^2-m_{K^0}^2)_{EM}+
(m_{K^+}^2-m_{K^0}^2)_{QM}
\end{equation}
Employing the value of $(m_{K^+}^2-m_{K^0}^2)_{EM}$ and experimental data of
mass difference between $K^+$ and $K^0$\cite{PDG}, we obtain
\begin{equation}
(m_{K^+}^2-m_{K^0}^2)_{QM}=-0.006346GeV^2=-2 m_K\times 6.4 MeV
\end{equation}
The use of the result of $(m_{K^+}^2-m_{K^0}^2)_{QM}$ together with
Eq.(17) will yield mass ratios of light quarks
\begin{eqnarray*}
& &\frac{m_u+m_d}{m_s+\hat{m}}=\frac{f_\pi^2 m_\pi^2}{f_k^2 m_K^2}=0.070, \\
& &\frac{m_d-m_u}{m_s-\hat{m}}=\frac{f_k^2(m_{K^0}^2-m_{K^+}^2)_{QM}}
{f_k^2 m_K^2-f_\pi^2 m_\pi^2}=0.028.
\end{eqnarray*}
where $\hat{m}=(m_u+m_d)/2$. These above results can be translated into
\begin{eqnarray*}
\frac{m_d}{m_s}=0.050,\;\;\;\frac{m_d-m_u}{m_s}=0.027,\;\;\;
\frac{m_u}{m_d}=0.44.
\end{eqnarray*}
The results are in agreement with the data of light quark mass ratios[19].
Similar results are recently given by Bijnens et al.\cite{BPR},
Leutwyler\cite{Le} and Duncan et al.\cite{DET}.
The value of $\frac{m_u}{m_d}=0.44$ reflects the breaking of isospin symmetry
in the present theory.

Finally, using the value of $m_s=175\pm16MeV$ which is obtained with
QCD sum rules\cite{JC}
in the $\overline{\mbox{MS}}$ scheme at scale $\mu=1GeV$,
we can calculated $m_u$ and $m_d$ with
the above mass ratios. The result reads
\begin{eqnarray*}
m_u(1GeV^2)=3.8\pm0.3MeV,\;\;\;\;m_d(1GeV^2)=8.7\pm0.8MeV,\\
\end{eqnarray*}

\section{$K_1^+-K_1^0$ electromagnetic mass difference}

The Lagrangians ${\cal L}_{K_1 K_1 v v}$, ${\cal L}_{K_1 K_1 v}$ and
${\cal L}_{K_1 K v}$ which contribute to electromagnetic self-energies of
$K_1$-meson are
\begin{eqnarray}
{\cal L}_{K_1 K_1 v v}&=&-\frac{2}{g^2}[
 \rho_\mu^3 v^{8\mu}(K_{1\nu}^+ K_1^{-\nu}-K_{1\nu}^0\bar{K}_1^{0\nu})\nonumber
 \\
& &-\frac{\gamma^2}{2}\rho_\mu^3 v_{\nu}^8(K_1^{+\mu} K_1^{-\nu}-K_1^{0\mu}
\bar{K}_1^{0\nu}+ h.c.)] \\
{\cal L}_{K_1 K_1 v}&=&\frac{i}{g}(1-\frac{\gamma^2}{\pi^2 g^2})[
\partial^\nu\rho_{\mu}^3(K_1^{+\mu}K_{1\nu}^{-}-K_{1}^{0\mu}\bar{K}_{1\nu}^{0})
+\partial^\nu v_\mu^8(K_1^{+\mu}K_{1\nu}^-+K_1^{0\mu}\bar{K}_{1\nu}^0)]
\nonumber \\
& &-\frac{i}{g}\rho_\nu^3[K_1^{+\mu}(\partial^\nu K_{1\mu}^--\gamma^2
\partial_{\mu}K_{1}^{-\nu})-K_1^{0\mu}(\partial^\nu\bar{K}_{1\mu}^0-
\gamma^2\partial_{\mu}\bar{K}_1^{0\nu})]
\nonumber \\
& &-\frac{i}{g}v_\nu^8[K_1^{+\mu}(\partial^\nu K_{1\mu}^--\gamma^2
\partial_{\mu}K_{1}^{-\nu})+K_1^{0\mu}(\partial^\nu\bar{K}_{1\mu}^0-
\gamma^2\partial_{\mu}\bar{K}_1^{0\nu})]
\nonumber \\
& & + h.c.  \\
{\cal L}_{K K_1 v}&=&\frac{i}{g}\beta_1[\rho_\mu^3(K^{+}K_1^{-\mu}-
K^0\bar{K}_1^{0\mu})+v_\mu^8(K^{+}K_1^{-\mu}+
K^0\bar{K}_1^{0\mu})]\nonumber \\
& &+\frac{i}{g}\beta_2[\rho_\nu^3(\partial^{\mu\nu} K^+K_{1\mu}^-
-\partial^{\mu\nu} K^0\bar{K}^0_{1\mu})+v_\nu^8
(\partial^{\mu\nu} K^{+}K_{1\mu}^-+\partial^{\mu\nu} K^0
\bar{K}^0_{1\mu})]\nonumber \\
& &+\frac{i}{g}\beta_3[\rho_\mu^3(K_1^{+\mu}\partial^2 K^{-}-
K_1^{0\mu}\partial^2 \bar{K}^0)+v_\mu^8
(K_1^{+\mu}\partial^2 K^{-}+K_1^{0\mu}\partial^2\bar{K}^0)]\nonumber \\
& &-\frac{i}{g}\beta_4[\rho_\mu^3(K^{+}\partial^2 K_1^{-\mu}-
K^{0}\partial^2 \bar{K}_1^{0\mu})+v_\mu^8
(K^{+}\partial^2 K_1^{-\mu}+K^{0}\partial^2\bar{K}_1^{0\mu})]\nonumber \\
& &-\frac{i}{g}\beta_5[\rho_\mu^3(\partial_\nu K_{1\mu}^+ \partial^\nu K^-
-\partial_\nu K_{1\mu}^0 \partial^\nu \bar{K}^0)+
v_\mu^8(\partial_\nu K_{1\mu}^+\partial^\nu K^-
+\partial_\nu K_{1\mu}^0 \partial^\nu \bar{K}^0)]\nonumber \\
& & + h.c.
\end{eqnarray}
The photon-mesons interaction Lagrangians can be obtained by combining the
above Lagrangians with substitutions (9),(10),(11),and the corresponding
Feynman diagrams have been shown in Figs.(11),(12),(13). The examination
of gauge-independence can be done in the same way as in the preceding Sections.

From Fig.(11),we have
\begin{eqnarray}
& &(m_{K_1^+}^2-m_{K_1^0}^2)_1\nonumber\\
&&=ie^2\frac{\gamma^2\langle K_1| \int d^4 x
K_1^{\mu+}K_1^{\nu-}|K_1\rangle-\langle K_1| \int d^4 x K_1^{\lambda+} K_{1\lambda}^-|K_1\rangle
g^{\mu\nu}}{\langle K_1| \int d^4 x K_{1\mu}^+K_1^{\mu-}|K_1\rangle}\nonumber \\
& &\int\frac{d^4 k}{(2\pi)^4}
(g_{\mu\nu}-\frac{k_\mu k_\nu}{k^2})
[\frac{1}{3}\frac{m_\rho^2 m_\omega^2}{k^2(k^2-m_\rho^2)(k^2-m_\omega^2)}+
\frac{2}{3}\frac{m_\rho^2 m_\phi^2}{k^2(k^2-m_\rho^2)(k^2-m_\phi^2)}]
\end{eqnarray}
From Fig.(12),we obtain
\begin{eqnarray}
& &(m_{K_1^+}^2-m_{K_1^0}^2)_2\nonumber \\
& &=\frac{ie^2}{\langle K_1| \int d^4 x K_1^{\mu+}
K_{1\mu}^-|K_1\rangle}\int\frac{d^4 k}{(2\pi)^4}\frac{1}{k^2-2p\cdot k}\nonumber \\
& &\{\langle K_1| \int d^4 x K_1^{\mu+}K_{1\mu}^-|K_1\rangle[4 m_{K_1}^2+(b^2+2b\gamma^2)k^2
+2\gamma^4 p\cdot k-\frac{4(p\cdot k)^2}{k^2}\nonumber \\
& &-\frac{1}{m_{K_1}^2}(b k^2-(b-\gamma^2)p\cdot k)^2]
+\langle K_1| \int d^4 x K_{1\mu}^+K_{1\nu}^-|K_1\rangle k^\mu k^\nu [
-(3b^2-4 b+4)\nonumber \\
& &+D(b+\gamma^2)^2+4\gamma^2
-6b\gamma^2-2\gamma^4-\frac{2\gamma^4 p\cdot k}{k^2}
+\frac{1}{m_{K_1}^2 k^2}(b k^2-2(1-\gamma^2)p\cdot k)^2]\}\nonumber \\
& &[\frac{1}{3}\frac{m_\rho^2 m_\omega^2}{k^2(k^2-m_\rho^2)(k^2-
m_\omega^2)}+\frac{2}{3}\frac{m_\rho^2 m_\phi^2}{k^2(k^2-m_\rho^2)(
k^2-m_\phi^2)}]
\end{eqnarray}
From Fig.(13),we get
\begin{eqnarray}
& &(m_{K_1^+}^2-m_{K_1^0}^2)_3\nonumber \\
& &=\frac{-ie^2}{\langle K_1| \int d^4 x K_{1\mu}^+
K_1^{\mu-}|K_1\rangle}\int\frac{d^4 k}{(2\pi)^4}\frac{1}{(p-k)^2-m_K^2}\nonumber\\
& &\{\langle K_1| \int d^4 x K_{1\mu}^+K_1^{\mu-}|K_1\rangle(\beta_1^\prime-3\beta_2 p\cdot k+
\beta_3 k^2)^2+\nonumber \\
& &\langle K_1| \int d^4 x K_{1\mu}^+K_{1\nu}^-|K_1\rangle k^\mu k^\nu
[\beta_2 m_{K_1}^2-\frac{(\beta_1^\prime-2 \beta_2 p\cdot k+\beta_3 k^2)^2}{k^2}
]\}\nonumber \\
& &[\frac{1}{3}\frac{m_\rho^2 m_\omega^2}{k^2(k^2-m_\rho^2)(k^2-
m_\omega^2)}+\frac{2}{3}\frac{m_\rho^2 m_\phi^2}{k^2(k^2-m_\rho^2)(
k^2-m_\phi^2)}]
\end{eqnarray}
with
$$
\beta_1^\prime=\beta_1+(\beta_3+\beta_4-\beta_5) m_{K_1}^2.
$$

Comparing Eqs.(87)-(89) and Eqs.(49)(51)(54) with taking $f_k=f_\pi$,
$m_K^2=m_\pi^2=0$,and $m_\rho=m_\omega=m_\phi$, $m_a=m_{K_1}$,
we can conclude that
\begin{eqnarray*}
(m_{a^+}^2-m_{a^0}^2)_i=(m_{K_1^+}^2-m_{K_1^0}^2)_i,\;\;\; i=1,2,3.
\end{eqnarray*}
This means that the square mass difference coming from electromagnetic
interaction between the charged axial-vector mesons and their corresponding
neural partners are equal in the chiral SU(3) limit, i.e.
\begin{equation}
(m_{a^+}^2-m_{a^0}^2)_{EM}=(m_{K_1^+}^2-m_{K_1^0}^2)_{EM}
\end{equation}
which is similar to Dashen's theorem for the pseudoscalar $\pi$ and $K$-mesons.
Certainly, the SU(3) symmetry breaking will bring about the violation of the
above equation.

After carrying out the Feynman integrations of Eqs.(87)-(89),
the numerical results for $m_{K_1^+}^2-m_{K_1^0}^2$ are
\begin{eqnarray*}
& &(m_{K_1^+}^2-m_{K_1^0}^2)_1=-0.000781GeV^2,\\
& &(m_{K_1^+}^2-m_{K_1^0}^2)_2=-0.003474GeV^2,\\
& &(m_{K_1^+}^2-m_{K_1^0}^2)_3=0.001252GeV^2.
\end{eqnarray*}
Thus, the correction of the electromagnetic mass to $K_1(1400)$ mesons is
\begin{equation}
(m_{K_1^+}^2-m_{K_1^0}^2)_{EM}=-0.003003GeV^2=-2 m_{K_1}\times1.1MeV
\end{equation}
This result gives a very large violation of Eq.(90).
\begin{eqnarray}
{(m_{K_1^+}^2-m_{K_1^0}^2)_{EM}\over(m_{a^+}^2-m_{a^0}^2)_{EM}}=2.08
\end{eqnarray}

\section{$K^{*+}$-$K^{*0}$ electromagnetic mass difference}

The Lagrangians contributing to $m_{K^{*+}}-m_{K^{*0}}$ come from
both the normal part of the effective Lagrangian ${\cal L}_{RE}$
and the abnormal part ${\cal L}_{IM}$. ${\cal L}_{K K^* v}$
deriving from ${\cal L}_{IM}$ is exactly Eq.(66).
\begin{eqnarray}
{\cal L}_{K^* K^* v v}&=&-\frac{2}{g^2}\rho_{\mu}^3 v^{8\mu}(K_\nu^+ K^{-\nu}-
K_\nu^0\bar{K}^{0\nu})\nonumber \\
& &+\frac{1}{g^2}\rho_\mu^3 v_\nu^8(K^{+\mu}K^{-\nu}-
K^{0\mu}\bar{K}^{0\nu}+h.c.),\\
{\cal L}_{K^* K^* v}&=&\frac{i}{g}[\partial_\nu \rho_\mu^3(K^{+\mu}K^{-\nu}-
K^{0\mu}\bar{K}^{0\nu})+\partial_\nu v_\mu^8(K^{+\mu}K^{-\nu}+
K^{0\mu}\bar{K}^{0\nu})]\nonumber \\
& &-\frac{i}{g}\rho_\nu^3[K_\mu^+(\partial^\nu K^{-\mu}-
\partial^\mu K^{-\nu})-K_\mu^0(\partial^\nu \bar{K}^{0\mu}-
\partial^\mu \bar{K}^{0\nu})]\nonumber\\
& &-\frac{i}{g}v_\nu^8[K_\mu^+(\partial^\nu K^{-\mu}-
\partial^\mu K^{-\nu})+K_\mu^0(\partial^\nu \bar{K}^{0\mu}-
\partial^\mu \bar{K}^{0\nu})]\nonumber \\
& & +h.c.
\end{eqnarray}

Substitutions (9),(10) and (11) together with Eq.(66),(93) and (94)
will produce the photon-$K^*$ mesons
interaction Lagrangian ${\cal L}_{K^* K^* \gamma \gamma}$,
${\cal L}_{K^* K^* \gamma}$ and ${\cal L}_{K^* K \gamma}$.
The one-loop Feynman diagrams contributing to electromagnetic mass splitting
of $K^{*+}$ and $K^{*0}$ are shown in Figs.(14),(15) and (16).
The gauge dependent terms from Figs.(14a) and (15a) will
vanish in the framework of dimensional regularization, and one from Fig.(16)
will also vanish due to the totally antisymmetric tensor
$\epsilon^{\mu\nu\alpha\beta}$ in Eq.(66). It is
straightforward to evaluate the contributions to $m_{K^*+}^2-m_{K^*0}^2$
from Figs.(14)-(16) one by one.

Contribution of Fig.(14) is
\begin{eqnarray}
& &(m_{K^{*+}}^2-m_{K^{*0}}^2)_1\nonumber \\
& &=-\frac{i9e^2}{4}\int\frac{d^4 q}{(2\pi)^4}[
\frac{1}{3}\frac{m_\rho^2 m_\omega^2}{q^2(q^2-m_\rho^2)(q^2-m_\omega^2)}+
\frac{2}{3}\frac{2 m_\rho^2 m_\phi^2}{q^2(q^2-m_\rho^2)(q^2-m_\phi^2)}]
\end{eqnarray}

Contribution of Fig.(15) is
\begin{eqnarray}
& &(m_{K^{*+}}^2-m_{K^{*0}}^2)_2 \nonumber \\
& &=\frac{ie^2}{\langle K^*| \int d^4 x K_\mu^{+}K^{-\mu}|K^*\rangle}
\int\frac{d^4 q}{(2\pi)^4}\{\langle K^*| \int d^4 x K_\mu^{+}K^{\mu-}|K^*\rangle\nonumber\\
& &\times[k^2-\frac{(k^2)^2}{m_{K^*}^2}+4p^2+4 q^2
+\frac{4(p\cdot q)^2}{m_{K^*}^2}-\frac{4 q^2 p\cdot q}{m_{K^*}^2}-
\frac{4 (p\cdot q)^2}{q^2}]+\nonumber \\
& &\langle K| \int d^4 x K_{\mu}^+K_{\nu}^-|K^*\rangle q^\mu q^\nu[8
+\frac{4p\cdot q}
{m_{K^*}^2}-\frac{4(p\cdot q)^2}{q^2 m_{K^*}^2}-\frac{k^2}{q^2}+
\frac{(k^2)^2}{q^2 m_{K^*}^2}]\}\nonumber \\
& &\times\frac{1}{k^2-m_{K^*}^2}[\frac{1}{3}\frac{m_\rho^2 m_\omega^2}{
q^2(q^2-m_\rho^2)(q^2-m_\omega^2)}+\frac{2}{3}\frac{m_\rho^2 m_\phi^2}
{q^2(q^2-m_\rho^2)(q^2-m_\phi^2)}]
\end{eqnarray}
where $p$ is the external momentum of $K^*$-mesons,$k=p-q$. For mass-shell $K^*$ mesons,
$p^2$=$m_{K^*}^2$,and $p^\mu K_\mu(p)=0$. Here $K_\mu(p)$ is the Fourier
transformation of $K^*$-mesons field
$$K_\mu(p)=\frac{1}{(2\pi)^4}\int d^4 x K_\mu(x) e^{-ipx}. $$

Contribution of Fig.(16) is
\begin{eqnarray}
& &(m_{K^{*+}}^2-m_{K^{*0}}^2)_3=\frac{ie^2}{\langle K^*| \int d^4 x K_\mu^{+} K^{-\mu}
|K^*\rangle}\frac{9}{4\pi^4 g^2 f_k^2}\int\frac{d^4 q}{(2\pi)^4}\nonumber \\
& &\{\langle K^*| \int d^4 x K_\mu^+ K^{-\mu}|K^*\rangle[p^2 q^2-(p\cdot q)^2]-
\langle K^*| \int d^4 x K_\mu^+ K_\nu^-|K^*\rangle q^{\mu} q^{\nu} p^2 \}
\nonumber \\
& &\frac{1}{k^2-m_K^2}[\frac{1}{3}\frac{m_\rho^2 m_\omega^2}{q^2(q^2-m_\rho^2)
(q^2-m_\omega^2)}-\frac{2}{3}\frac{m_\rho^2 m_\phi^2}{q^2(q^2-m_\rho^2)
(q^2-m_\phi^2)}]
\end{eqnarray}

The Feynman integrations of $(m_{K^{*+}}^2-m_{K^{*0}}^2)_{1,3}$ are
finite, only the logarithmic divergence emerges in  $(m_{K^{*+}}^2-
m_{K^{*0}}^2)_2$, which can be factorized by using Eq.(37). The performing of these
Feynman integrations is standard. The numerical results are
\begin{eqnarray*}
& &(m_{K^{*+}}^2-m_{K^{*0}}^2)_1=-0.000938GeV^2,\\
& &(m_{K^{*+}}^2-m_{K^{*0}}^2)_2=-0.001547GeV^2,\\
& &(m_{K^{*+}}^2-m_{K^{*0}}^2)_3=-0.000662GeV^2.
\end{eqnarray*}
The electromagnetic mass correction to $K^*(892)$-mesons totally is
\begin{equation}
(m_{K^{*+}}^2-m_{K^{*0}}^2)_{EM}=
-0.003147 GeV^2=-2 m_{K^*}\times 1.76 MeV
\end{equation}
However, mass difference between $K^{*+}$ and $K^{*0}$ doesn't only
receive contributions from the virtual photon-exchange,but also from
the other nonelectromagnetic interactions, such as isospin symmetry breaking,
which is similar to the case of pseudoscalar-$K$ mesons.
So we have
\begin{eqnarray}
(m_{K^{*+}}-m_{K^{*0}})_{EXPT}=(m_{K^{*+}}-m_{K^{*0}})_{EM}+
(m_{K^{*+}}-m_{K^{*0}})_{nonEM}
\end{eqnarray}
Using the experimental value of $(m_{K^{*+}}-m_{K^{*0}})_{EXPT}=
-6.7\pm 1.2MeV$\cite{PDG}, we obtain
\begin{equation}
(m_{K^{*+}}-m_{K^{*0}})_{nonEM}=-4.94\pm1.2MeV
\end{equation}

The nonelectromagnetic mass difference of $(m_{K^{*+}}-m_{K^{*0}})_{nonEM}$
,which comes from isospin breaking effects, has ever been evaluated\cite{SSW,DGH}.
In Ref.\cite{SSW}, J.Schechter et al. predicted that $(m_{K^{*+}}-m_{K^{*0}})_{nonEM}$ would be from $-2.04MeV$
to $-6.78MeV$. By choosing the best fitted parameter,
they concluded that $(m_{K^{*+}}-m_{K^{*0}})_{nonEM}$=$-4.47MeV$,
which is close to our result of Eq.(100).

\section{Summary and Discussions}
In the framework of the present theory, the dynamics of meson-fields
comes from the quark-loop integrations within mesonic background fields.
The logarithmic divergence due
to the quark-loop integrations is absorbed into the
parameter $g$ (Eq.(3)) in this truncated
field theory. Thus, both meson's effective Lagrangians with VMD
and criterion to factorize the logarithmic divergences in the loop calculations
are well established. In this paper, by using this theory, we have computed
all one-loop diagrams contributing to the electromagnetic mass splitting of the
low-lying mesons including pseudoscalar mesons $\pi$ and $K$,
axial-vector mesons $a_1 $ and $K_1(1400)$, and vector meson $K^*(892)$.
Fortunately, no other higher order divergences but the logarithmic
divergences are emerging in the Feynman integrations of the above loop
diagrams.
Therefore it is reasonable to factorize these logarithmic divergences by using
the intrinsic parameter $g$ in this theory, which is determined by the
experimental values of $f_\pi$,$m_\rho$ and $m_{a_1}$. Then,
it is unnecessary to introduce other additional
parameters or counterterms into this theory to absorb the mesonic loop
divergences.
The dimensional regularization has been employed and the
gauge-independence of the calculations is examined.

The electromagnetic mass splittings of $\pi$ and $a_1$ are calculated in
the chiral limit because of the smallness of $u$ and $d$ quark masses,
and the result of $m_{\pi^+}-m_{\pi^0}$ is close to the experimental data.
However, the electromagnetic mass splittings of the strange-flavor mesons
$K$,$K_1$ and $K^*$ have been evaluated to the order of $m_s$ or $m_K^2$
because of the large strange quark mass.
Thus, a rather large violation of Dashen's theorem (which holds in the
chiral SU(3) limit of the present theory)has been revealed at the leading order
in quark mass expansion. The mass ratios of light quarks
has been calculated,and masses of $u$, $d$ quarks have been estimated by employing the
value of $m_s$ obtained with QCD sum rules.
It has been found that there exists a new relation
for axial-vector mesons, i.e. $(m_{a^+}^2-
m_{a^0}^2)_{EM}=(m_{K_1^+}^2-m_{K_1^0}^2)_{EM}$ is obeyed in the chiral SU(3)
limit. Moreover, the non-electromagnetic mass difference between $K^{*+}$ and
$K^{*0}$ is estimated by using the experimental value of $(m_{K^{*+}}-m_{K^{*0}})$
with $(m_{K^{*+}}-m_{K^{*0}})_{EM}$ calculated in this paper.

The electromagnetic self-energies of the other low-lying mesons,
such as  vector mesons $\rho$,$\omega$,$\phi(1020)$,and
pseudoscalar mesons $\eta$,$\eta^{\prime}(960)$,also
need to be evaluated. However, the quadratic or more high order
divergences will emerge in the Feynman integrations of the loop calculations
of $\rho$,$\omega$,and $\phi$. It is unsuitable to factorize
these higher order divergences
by the parameter $g$ in which only the logarithmic divergence is involved.
As for $\eta$ and $\eta^{\prime}$, U(1) anomaly problem and
the mixing of $\eta$-$\eta^{\prime}$ should be taken into account.
The investigation on these problems is beyond the scope of the present work.

\begin{center}
{\bf ACKNOWLEDGMENTS}
\end{center}
D.N.Gao and M.L.Yan are partially supported by NSF of China through
C.N.Yang. B.A.Li is partially supported by DOE Grant No. DE-91ER75661.
\vskip1.0cm
\centerline{\bf Appendix A: Feynman Rules and the Photon
Propagator Within $\rho$}
\vskip0.8cm

1, The propagators taken in this paper are as following:

Pseudoscalar-meson fields,
\begin{eqnarray*}
\lefteqn{
\langle 0|T \phi(x) \phi(y)
|0\rangle =\dk  \Delta_F(k)
e^{-ik(x-y)},}  \nonumber \\
&& \Delta_F(k)={i \over k^2-m^2+i\epsilon }\;\;\;\;\;\;\;\;\;\;\;(A1)
\end{eqnarray*}

Vector-meson fields,
\begin{eqnarray*}
\lefteqn{
\langle 0|T(V_\mu^i(x) V_\nu^j(y) 
|0\rangle =\dk \delta_{ij} 
\Delta_{F\mu\nu}(k)
e^{-ik(x-y)},}  \nonumber \\
&& \Delta_{F\mu\nu}(k)={-i \over 
k^2-m_V^2+i\epsilon }(g_{\mu\nu}-{k_\mu k_\nu \over m_V^2})
\;\;\;\;\;\;\;\;\;(A2)
\end{eqnarray*}
where $V_\mu^i(x)= a_\mu^i(x),\rho_\mu^i(x)$,$\omega_\mu (x)$,$\phi_{\mu}(x)$,
$K_{1\mu}(x)$ and $K_\mu(x)$.

2, The photon propagator within $\rho$: From Eq.(22) and (12),
we have
\begin{eqnarray*}
\lefteqn{
\Delta_{F\mu\nu}^{(\gamma\rho)}(x-y)=
\langle 0|T\{ A_\mu (x)A_\nu (y)
} \\
&&
-2i\dy A_\mu (x)\rho_\nu^3(y)
[\p_\lambda \rho_\sigma^3 (x_1)
(\p^\lambda A^\sigma (x_1)-\p^\sigma A^\lambda (x_1) )] \\
&&-{{1 \over 2} \ddx \rho_\mu^{\rm 3}}(x) \rho_\nu^3(y)
[\p_\lambda \rho_\sigma^3 (x_1)
(\p^\lambda A^\sigma (x_1)-\p^\sigma A^\lambda (x_1) )]  \\
&&[\p_\alpha \rho_\beta^3 (x_2)
(\p^\alpha A^\beta (x_2)-\p^\alpha A^\beta (x_2) )]
\}|0\rangle. \;\;\;\;\;\;\;\;\;\;\;\;\;\;(A3)
\end{eqnarray*}
Using Eq.(22) and Eq.(A2), we get
\begin{eqnarray*}
\Delta_{F\mu\nu}^{(\gamma\rho)}(x-y)&=&\int\frac{d^4 k}{(2\pi)^4}
(-i)\frac{1}{k^2} e^{-ik(x-y)}\{a\frac{k_\mu k_\nu}{k^2}+
(g_{\mu\nu}-\frac{k_\mu k_\nu}{k^2})\\
& &\times[1-\frac{2k^2}{k^2-m_\rho^2}+\frac{k^4}{(k^2-m_\rho^2)^2}]\}.
\end{eqnarray*}

Then
$$
{\Delta_{F\mu\nu}^{(\gamma\rho)}(x-y)={\dk (-i)}{1 \over k^2}
[{m_\rho^4 \over (k^2-m_\rho^2)^2}(g_{\mu\nu}-{k_\mu k_\nu
 \over k^2}) +a {k_\mu k_\nu \over k^2}]}
e^{-ik(x-y)}.\;\;\;\;\;\;\; (A4)
$$
This is Eq.(23).

\vspace{1.0cm}
\centerline{\bf Appendix B: $\Delta_{F_{1\mu\nu}}^{(\gamma v)}$ and
$\Delta_{F_{2\mu\nu}}^{(\gamma v)}$}
\vskip0.8cm

The photon propagator within $\rho$ can be generalized to the
photon propagator within $v$ including $\rho$,$\omega$ and
$\phi$ to simplify the corresponding calculations of the
strange-flavor mesons. In this Appendix, as an example, we
display the whole calculating process of Fig.(7) to deduce
$\Delta_{F_{1\mu\nu}}^{(\gamma v)}$.

${\cal L}_{K K v v}$ has been shown in Eq.(61). The corresponding photon-
mesons couplings ${\cal L}_{K K \gamma \gamma}$,${\cal L}_{K K \rho \gamma}$
${\cal L}_{K K\omega \gamma}$ and ${\cal L}_{K K \phi \gamma}$,
which contribute to electromagnetic mass differences
between $K^+$ and $K^0$, can be obtained by substitutions (9),(10) and (11).
All the one-loop Feynman diagrams contributing to $(m_{K^+}^2-m_{K^0}^2)_1$
are shown in Fig.(7a),(7b),(7c), and the corresponding $S$-matrices
are denoted as $S_K(1)_i$, $i=a,b,c$. Thus we have
\begin{eqnarray*}
S_K(1)_a&=&i\langle K| T\int d^4 x {\cal L}_{K K \gamma \gamma}(x)|K\rangle \nonumber \\
&=&\langle K| \int d^4 x K^+ K^-|K\rangle\frac{ie^2}{f_k^2}\frac{d^4 k}{(2\pi)^4}(
F_K^2+\frac{k^2}{2\pi^2}) \\
& &\frac{i}{-k^2}g^{\mu \nu}[(g_{\mu\nu}-\frac{k_\mu k_\nu}{k^2})
+a\frac{k_\mu k_\nu}{k^2}].\;\;\;\;\;\;\;\;\;\;\;\;\;\;\;\;\;\;\;\;\;(B1)
\end{eqnarray*}
and
\begin{eqnarray*}
S_K(1)_b&=&S_K(1)_{\rho}+S_K(1)_\omega+S_K(1)_\phi,
\end{eqnarray*}
with
\begin{eqnarray*}
S_K(1)_{\rho}&=&\frac{i^2}{2!}2\langle K| T\int d^4 x d^4 y {\cal L}_{K K \rho \gamma}(x)
{\cal L}_{\rho \gamma}(y)|K\rangle  \\
&=&\langle K| \int d^4 x K^+ K^-|K\rangle\frac{ie^2}{f_k^2}\frac{d^4 k}{(2\pi)^4}(
F_K^2+\frac{k^2}{2\pi^2}) \\
& &\frac{i}{(-k^2+m_\rho^2)(-k^2)}k^2 g^{\mu \nu}(g_{\mu\nu}-\frac{k_\mu k_\nu}{k^2})
,\;\;\;\;\;\;\;\;\;\;\;\;\;\;\;\;\;(B2)\\
S_K(1)_{\omega}&=&\frac{i^2}{2!}2\langle K| T\int d^4 x d^4 y {\cal L}_{K K \omega \gamma}(x)
{\cal L}_{\omega \gamma}(y)|K\rangle  \\
&=&\langle K| \int d^4 x K^+ K^-|K\rangle\frac{1}{3}\frac{ie^2}{f_k^2}\frac{d^4 k}{(2\pi)^4}(
F_K^2+\frac{k^2}{2\pi^2}) \\
& &\frac{i}{(-k^2+m_\omega^2)(-k^2)}k^2 g^{\mu \nu}(g_{\mu\nu}-\frac{k_\mu k_\nu}{k^2})
,\;\;\;\;\;\;\;\;\;\;\;\;\;\;\;\;(B3)\\
S_K(1)_{\phi}&=&\frac{i^2}{2!}2\langle K| T\int d^4 x d^4 y {\cal L}_{K K \phi \gamma}(x)
{\cal L}_{\phi \gamma}(y)|K\rangle \\
&=&\langle K| \int d^4 x K^+ K^-|K\rangle\frac{2}{3}\frac{ie^2}{f_k^2}\frac{d^4 k}{(2\pi)^4}(
F_K^2+\frac{k^2}{2\pi^2}) \\
& &\frac{i}{(-k^2+m_\phi^2)(-k^2)}k^2 g^{\mu \nu}(g_{\mu\nu}-\frac{k_\mu k_\nu}{k^2})
.\;\;\;\;\;\;\;\;\;\;\;\;\;\;\;\;(B4)
\end{eqnarray*}
and
\begin{eqnarray*}
S_K(1)_c&=&S_K(1)_{\rho\omega}+S_K(1)_{\rho\phi},
\end{eqnarray*}
with
\begin{eqnarray*}
S_K(1)_{\rho\omega}&=&\frac{i^3}{3!}6\langle K| T\int d^4 x d^4 y d^4 z{\cal L}_{K K \rho \omega}(x)
{\cal L}_{\rho \gamma}(y){\cal L}_{\omega \gamma}(z)|K\rangle  \\
&=&\langle K| \int d^4 x K^+ K^-|K\rangle\frac{1}{3}\frac{ie^2}{f_k^2}\frac{d^4 k}{(2\pi)^4}(
F_K^2+\frac{k^2}{2\pi^2})\\
& &\frac{i}{(-k^2+m_\rho^2)(-k^2+m_\omega^2)(-k^2)}(k^2)^2 g^{\mu \nu}(g_{\mu\nu}-\frac{k_\mu k_\nu}{k^2})
,\;\;\;\;\;\;\;\;(B5)\\
S_K(1)_{\rho\phi}&=&\frac{i^3}{3!}6\langle K| T\int d^4 x d^4 y d^4 z{\cal L}_{K K \rho \phi}(x)
{\cal L}_{\rho \gamma}(y){\cal L}_{\phi \gamma}(z)|K\rangle  \\
&=&\langle K| \int d^4 x K^+ K^-|K\rangle\frac{2}{3}\frac{ie^2}{f_k^2}\frac{d^4 k}{(2\pi)^4}(
F_K^2+\frac{k^2}{2\pi^2}) \\
& &\frac{i}{(-k^2+m_\rho^2)(-k^2+m_\phi^2)(-k^2)}(k^2)^2 g^{\mu \nu}
(g_{\mu\nu}-\frac{k_\mu k_\nu}{k^2}).\;\;\;\;\;\;\;(B6)
\end{eqnarray*}
Thus,the total contribution of Fig.(7) is
\begin{eqnarray*}
S_K(1)&=&\langle K| \int d^4 x K^+ K^-|K\rangle\frac{ie^2}{f_k^2}\int\frac{d^4 k}{(2\pi)^4}
(F_K^2+\frac{k^2}{2\pi^2})\frac{i}{-k^2}g^{\mu\nu} \\
& &[(\frac{1}{3}\frac{m_\rho^2 m_\omega^2}{(k^2-m_\rho^2)(k^2-m_\omega^2)}+
\frac{2}{3}\frac{m_\rho^2 m_\phi^2}{(k^2-m_\rho^2)(k^2-m_\phi^2))}(
g_{\mu\nu}-\frac{k_\mu k_\nu}{k^2})+a\frac{k_\mu k_\nu}{k^2}] \\
&=&\langle K|\int d^4 x K^+ K^-|K\rangle\frac{ie^2}{f_k^2}\int\frac{d^4 k}{(2\pi)^4}
(F_K^2+\frac{k^2}{2\pi^2})g^{\mu\nu}\Delta_{F_{1\mu\nu}}^{(\gamma v)}(k)
\;\;\;\;\;\;\;(B7)
\end{eqnarray*}
Here, $\Delta_{F_{1\mu\nu}}^{(\gamma v)}(k)$ is exactly Eq.(68).

Similar procedure can be easily applied to Figs.(8),(9),and(10). We can conclude
that Figs.(7),(8),(9),which receive contributions from the normal part of the
effective Lagrangian ${\cal L}_{RE}$, yield the same expression  of
$\Delta_{F_{1\mu\nu}}^{(\gamma v)}(k)$, however, Fig.(10), which is
from the abnormal part of the effective Lagrangian ${\cal L}_{IM}$,
gives the form of $\Delta_{F_{2\mu\nu}}^{(\gamma v)}(k)$, i.e. Eq.(69).
The difference between $\Delta_{F_{1\mu\nu}}^{(\gamma v)}(k)$ and
$\Delta_{F_{2\mu\nu}}^{(\gamma v)}(k)$ comes from that $\omega$ and $\phi$ mesons
fields are always appear as the combination $\omega_\mu-\sqrt{2}\phi_\mu$ in
${\cal L}_{RE}$, but as the combination $\omega_\mu+\sqrt{2}\phi_\mu$ in
${\cal L}_{IM}$(to see Eq.(66)).


\vskip1.0cm
\leftline{\bf Caption}
\begin{description}
\item[Figs.1,2,3] one-loop Feynman diagrams contributing to
electromagnetic mass difference between $\pi^+$ and $\pi^0$,
the curly line is the photon-line.
\item[Figs.4,5,6] one-loop Feynman diagrams contributing to
electromagnetic mass difference between $a_1^+$ and $a_1^0$,
the curly line is the photon-line.
\item[Figs.7,8,9,10] one-loop Feynman diagrams contributing to
electromagnetic mass difference between $K^+$ and $K^0$,
the curly line is the photon-line, $v$ denotes neutral vector mesons
 $\rho$,$\omega$ and $\phi$.
\item[Figs.11,12,13] one-loop Feynman diagrams contributing to
electromagnetic mass difference between $K_1^+$ and $K_1^0$,
the curly line is the photon-line, $v$ denotes neutral vector mesons
 $\rho$,$\omega$ and $\phi$.
\item[Figs.14,15,16] one-loop Feynman diagrams contributing to
electromagnetic mass difference between $K^{*+}$ and $K^{*0}$,
the curly line is the photon-line, $v$ denotes neutral vector mesons
 $\rho$,$\omega$ and $\phi$.
\end{description}
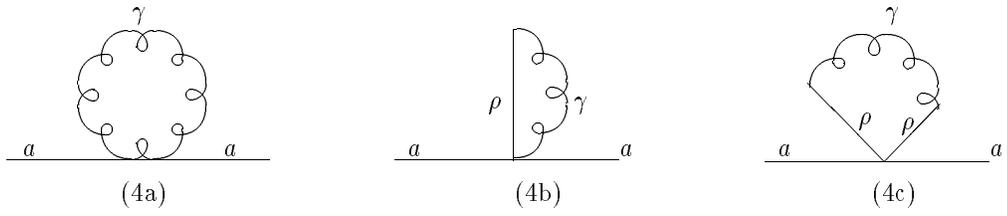
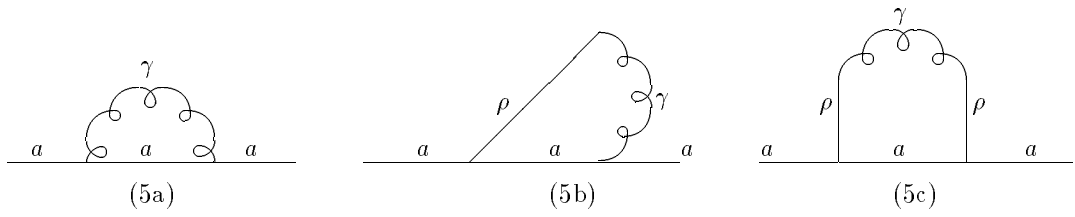
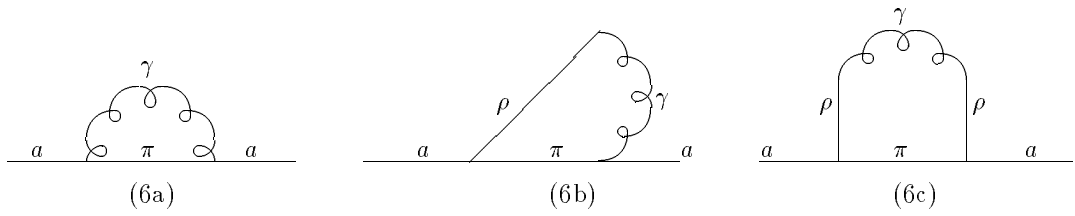
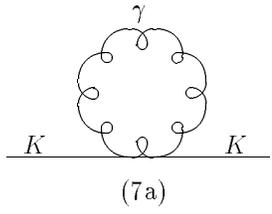
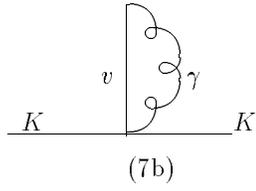
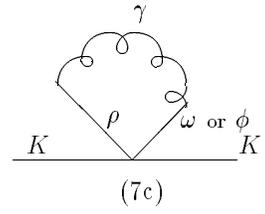
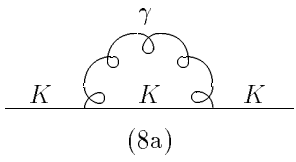
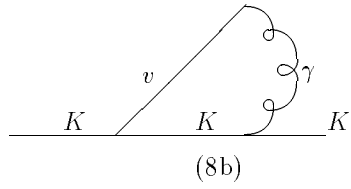
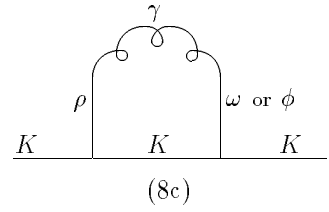
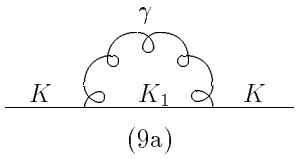
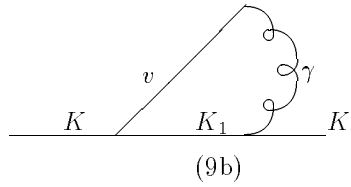
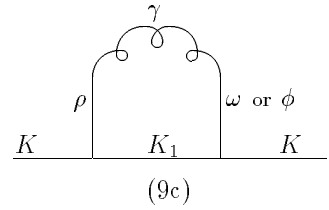
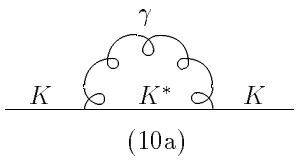
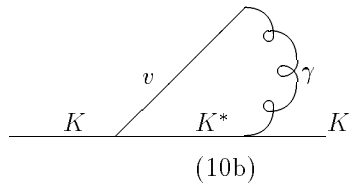
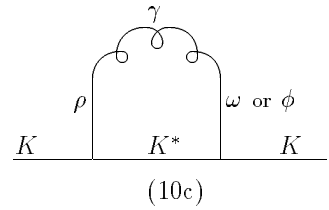
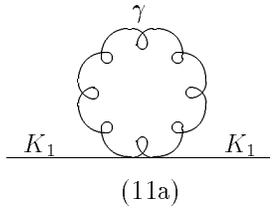
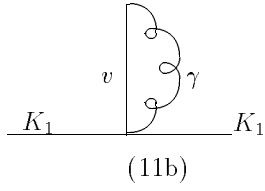
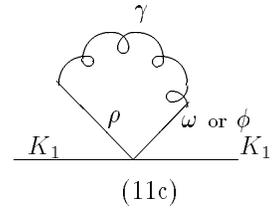
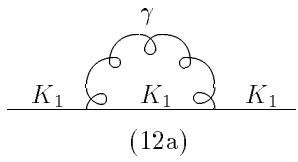
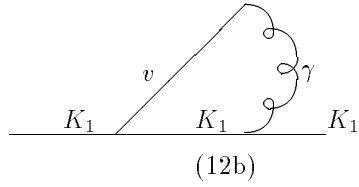
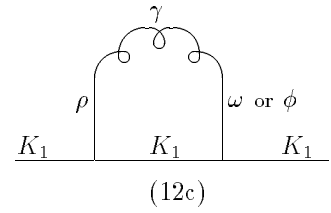
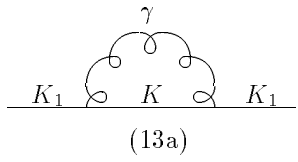
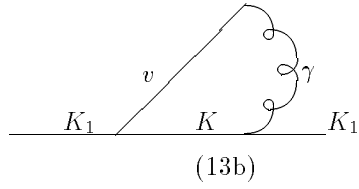
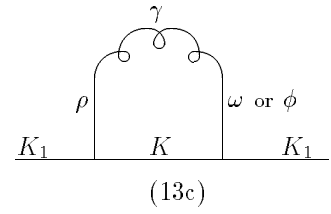
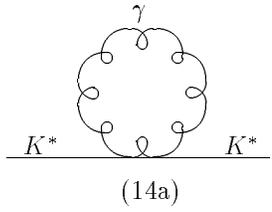
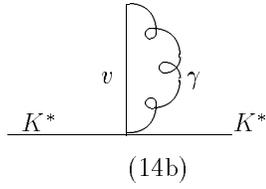
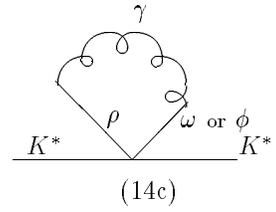
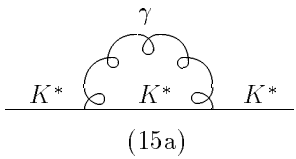
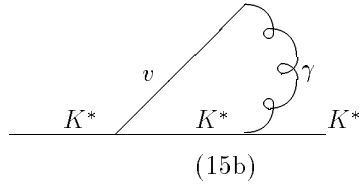
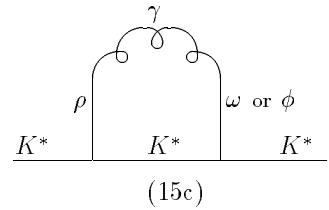
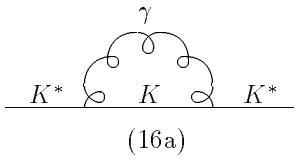
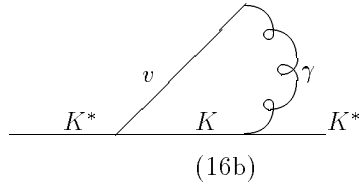
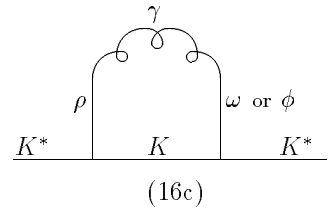
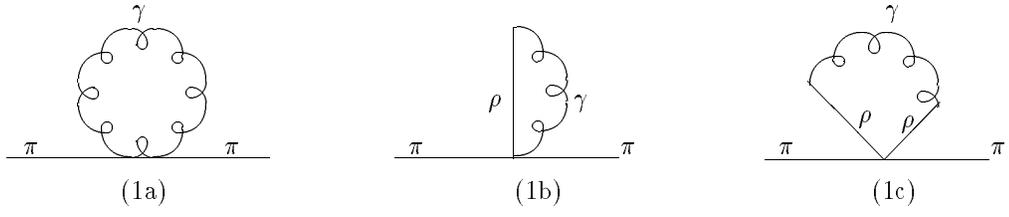
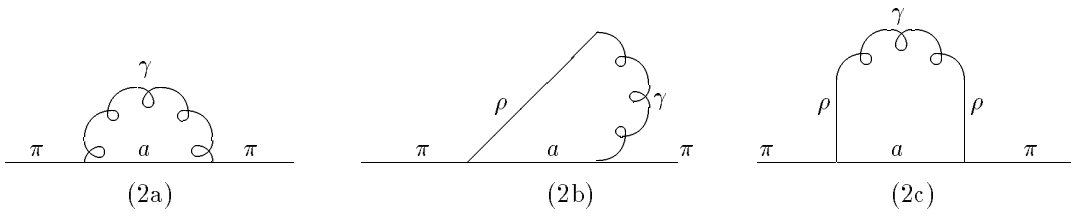
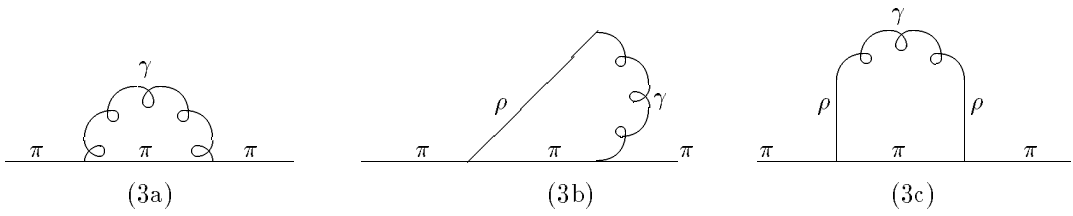
\end{document}